\documentstyle[graphicx, latin1]{mn}

\title{Failed disk winds; a physical origin for the soft X-ray excess?}

\author[N.J.\,Schurch \& C.\,Done]
{N.J.\,Schurch$^{*}$ \& C.\,Done\\ 
Department of Physics, Durham University, South Road, Durham, DH1 3LE, UK \\
$^{*}$ nicholas.schurch@durham.ac.uk}

\date{}

\def\asca{{\it ASCA\/}}

\def\xte{{\it RXTE\/}}
\def\xmm{{\it XMM-Newton\/}}

\def\Msun{\hbox{$\rm ~M_{\odot}$}}

\def\H0{{\rm ~km~s^{-1}~Mpc^{-1}}}

\def\arcs{{\hbox{$^{\prime\prime}$}}}

\def\etal{et al.~\/}

\def\la{\mathrel{\hbox{\rlap{\hbox{\lower4pt\hbox{$\sim$}}}{\raise2pt\hbox{$<$}}}}}
\def\ga{\mathrel{\hbox{\rlap{\hbox{\lower4pt\hbox{$\sim$}}}{\raise2pt\hbox{$>$}}}}}
\def\ls{\mathrel{\hbox{\rlap{\hbox{\lower4pt\hbox{$\sim$}}}\hbox{$<$}}}}
\def\gs{\mathrel{\hbox{\rlap{\hbox{\lower4pt\hbox{$\sim$}}}\hbox{$>$}}}}
\def\d25{D$_{\rm 25}$}

\def\.25{0.25 keV\thinspace}

\def\Mdotedd{\hbox{$\dot M_{Edd}$}}

\begin{document}

\maketitle 

\begin{abstract}

The origin of the soft X-ray excess emission observed in many type-1 AGN has been an unresolved problem in X-ray astronomy for over two decades. We develop the model proposed Gierli\'nski \& Done (2004), which models the soft excess with heavily smeared, ionized, absorption, by including the emission that {\it must} be associated with this absorption. We show that, rather than hindering the ionized absorption model, the addition of the emission actually helps this model reproduce the soft excess. The emission fills in some of the absorption trough, while preserving the sharp rise at $\sim$1 keV, allowing the total model to reproduce the soft excess curvature from a considerably wider range of model parameters. We demonstrate that this model is capable of reproducing even the strongest soft X-ray excesses by fitting it to the \xmm~EPIC PN spectrum of PG1211+143, with good results. The addition of the emission reduces the column density required to fit these data by a factor $\sim$2 and reduces the smearing velocity from $\sim$0.28c to $\sim$0.2c. Gierli\'nski \& Done suggested a tentative origin for the absorption in the innermost, accelerating, region of an accretion disk wind, and we highlight the advantages of this interpretation in comparison to accretion disk reflection models of the soft excess. Associating this material with a wind off the accretion disk results in several separate problems however, namely, the radial nature, and the massive implied mass-loss rate, of the wind. We propose an origin in a `failed wind', where the central X-ray source is strong enough to over-ionize the wind, removing the acceleration through line absorption before the material reaches escape velocity, allowing the material to fall back to the disk at larger radii.

\end{abstract}


\section{Introduction}
\label{1}

The origin of the soft X-ray ($\leq$2 keV) excess emission observed in many type-1 AGN has remained a long standing problem in X-ray astronomy despite considerable attention over the past two decades ({\it e.g.} Arnaud \etal 1985, Saxton \etal 1993, Mineshige \etal 2000). Several different ideas have been proposed for its origin including the thermal tail from the accretion disk ({\it e.g.} Pounds \etal 1986) and a separate cool Comptonizing region ({\it e.g.} Zdziarski \etal 1996) in either an optically thick transition region between the disk and a hot inner flow (Magdziarz \etal 1998) or in an ionized skin on the accretion disk (Janiuk, Czerny \& Madejski 2001). These models encounter considerable problems; specifically, that the best-fit temperature for such thermal components is remarkably constant across a wide range of objects that exhibit a considerable range of black hole masses, luminosities and accretion rates ({\it e.g.} Czerny \etal 2003; Gierli\'nski \& Done 2004: hereafter GD04; Crummy \etal 2005).

Detailed modelling of the \xmm~EPIC PN spectrum of the Narrow-line Seyfert 1 galaxy 1H 0707-495 suggested that Compton reflection from a partially ionized skin on the accretion disk can reproduce the strong soft excess (Fabian \etal 2002). This model provides a natural explanation for the `constant temperature' of the soft excess emission since the `excess' is now primarily associated with the abrupt change in opacity at $\sim$0.7 keV. Partially ionized material is much more reflective below this energy, and the rising reflected continuum is enhanced by associated CNO lines and radiative recombination (RRC) emission features. Together, these can produce the soft excess, with the advantage that the characteristic energy of the emission is fixed by atomic physics, so is largely independent of the properties of the AGN (Zycki \& Czerny 1994). The narrow atomic features associated with ionized reflection are not evident in the spectra of the soft excess suggesting that large (relativistic) velocity smearing is present in this material (Fabian \etal 2002). In addition, the largest soft excesses need a reflected fraction considerably greater than unity, arguing for an unusual geometry such as {\it i}) a highly clumpy/corrugated disk surface so that the observed spectrum is dominated by multiple reflections of the mostly hidden intrinsic hard X-ray source (Fabian \etal 2004; Ross \etal 2002), or {\it ii}) a model in which strong light-bending towards the black hole increases the X-ray flux incident on the disk ({\it e.g.} Fabian \etal 2005). The first of these models implies a complex physical configuration, perhaps as the result of disk instabilities that may be produced in high mass accretion rate sources such as Narrow-line Seyfert 1s (NLS1s), while the latter naturally produces the required large velocity smearing from the inner regions of the accretion disk.

Gierli\'nski \& Done (2004) proposed an alternative origin for the soft excess originating from similar atomic processes in partially ionized material, but with a different geometry. They suggested that the partially ionized material is seen in absorption, rather than in reflection. Again, a large velocity dispersion is required to hide the atomic line and edge features, and results in a broad, smooth absorption trough that adds considerable curvature to the X-ray spectrum between $\sim$0.3-5 keV. The advantage of this model over reflection is that a larger range soft excess sizes can be produced without invoking geometric changes. However, while the reflection models include the self-consistent emission from the partially ionized material, the absorption model presented in GD04 does not. This is equivalent to assuming that the material subtends a small solid angle, in contrast with the large fraction of type-1 AGN that show soft excess emission. Chevallier \etal (2005) include this emission as well as reflection and suggest that all three components work together to produce the soft X-ray excess.

GD04 suggest a tentative origin for the absorbing material in a UV line-driven accretion disk wind; that is material launched from the disk that is subsequently accelerated to high velocities by the radiation pressure on UV absorption lines, in similar fashion to the winds from hot stars ({\it e.g.} Lamers \& Cassinelli 1999), reasoning that a large velocity dispersion is implicit to material launched from the inner disk and hence strong velocity smearing is to be expected. The physical properties of such line-driven accretion disk winds have been investigated in considerable detail in other work ({\it e.g.} Arav, Li \& Begelman 1994, Murray \& Chiang 1995, 1997 \& 1998, Murray \etal 1995, Proga, Stone \& Kallmann 2000 - hereafter PSK00, Proga \& Kallman 2004 - hereafter PK04, etc). The general properties implied for accretion disk winds match well to the blue-shifted ($\leq$10$^{3}$ kms$^{-1}$) optical-UV-X-ray absorption lines seen in the Broad Absorption Line (BAL) quasars (Proga \& Kallman 2004), and may also be responsible for the broad emission lines seen in all Quasars (Murray \& Chiang 1997; Elvis 2000). Furthermore, such a wind provides an attractive and simple way of dynamically linking the otherwise distinct regions within the standard AGN unification scheme.

In this paper we extend the absorption model proposed in GD04 to include the emission that {\it must} be associated with this absorption. In Section~\ref{2} we briefly present the details of the photoionization modelling and the velocity smearing. In Section~\ref{3} we examine the properties of the absorption and emission in more detail, focusing on the impact of the emission on the resulting spectral shape, and demonstrate that the combination of emission and absorption removes many of the fine-tuning issues present in the pure absorption model of GD04.

We test the impact of the addition of the emission on fits to real data by fitting the \xmm~spectrum of the bright, strong soft excess Quasar PG1211+143 with our models of the absorption and emission from a disk wind, together with ionized reflection and the narrow absorption systems known to be present in this source (Section~\ref{4}). Sections 5 and 6 discuss the physical origin of the absorbing material and its properties.

\section{Modelling the wind}
\label{2}

\subsection{Photoionized material}
\label{2-1}

We calculate the absorption and emission from photoionized material using the atomic code XSTAR\footnote{http://heasarc.gsfc.nasa.gov/docs/software/xstar/xstar.html} v2.1kn3 (Bautista \& Kallman 2001). Given that we associate the smeared, blurred, ionized absorption with emission and absorption in the inner, accelerating, regions of a line-driven wind from the accretion disk, we chose gas conditions for the XSTAR model based in part on the values the data require and part on the simulations of AGN accretion disk winds presented in PSK00 \& PK04. Specifically, the data require a column density range of N$_{H}\sim$10$^{23-24}$ cm$^{-2}$ (GD04) while the simulations suggest a range of ionization states and a density of log$_{10}$($\xi$)=1.5-3.5 \& n=10$^{12}$ cm$^{-3}$, respectively, for the inner regions of the wind. XSTAR self-
consistently calculates the temperature of the material through the wind, we use solar metal abundances and we include the effects of mild turbulence in the gas (with a velocity dispersion of 100 km s$^{-1}$). The parameters defining the ionizing continuum were chosen to best approximate the properties of the emission from the inner regions of a NLS1 quasar. Specifically, the energy spectrum of ionizing X-rays is described by a power-law ({\it i.e.} N(E)=$K_{pl}$E$^{-\Gamma}$) with a slope of $\Gamma$=2.4 (an energy spectral index of -1.4 in XSTAR) and an intrinsic luminosity of L$_{int}$=10$^{46}$ erg s$^{-1}$ between 1-1000 Rydbergs.

Absorption seen along a line-of-sight is (to first order) independent of the covering fraction of the absorbing material, however the relative strength of the emission associated with this absorbing material {\it is} sensitive to the covering fraction. We account for this fact later in this analysis and, initially, calculate both the absorption and emission spectra for a covering fraction corresponding to material covering the entire sky from the point of view of the continuum source. A more detailed description of the treatment of covering fraction is given in Section~\ref{3-3}.

There is considerable discussion in the community concerning the physical conditions under which photoionization calculations such as these should be calculated. Specifically, whether one can assume either constant density or constant pressure throughout the material. There is no clear consensus on which assumption best represents the real physical conditions. Chevallier \etal (2005) argue for constant pressure conditions and highlight one of the more serious problems with the constant density model presented in GD04, namely that this model is extremely sensitive to the free parameters, particularly the ionization state, and that considerable fine tuning of these parameters is required in order to to match the observations of the soft X-ray excess. The constant pressure equilibrium models of Chevallier \etal avoid much of the fine tuning problem because the illuminating X-rays naturally induce a thermal instability within the material (e.g. Krolik, McKee \& Tarter 1981). Beyond this, the material may fragment into small, cold, dense clumps with low covering factor (R\'o$\rm \dot z$a\'nska \etal 2002, R\'o$\rm \dot z$a\'nska \etal 2005, Chevallier \etal 2005). Thus the instability effectively leads to disruption of the lower ionization portion of the cloud, generating absorbed spectra with a similar overall shape and, crucially, a similar absorption trough depth for a wide range of ionization states and column density.

Despite this advantage we choose to perform the photoionization calculations under the assumption of constant density primarily because the physical parameters are somewhat simpler and more intuitive. In Section~\ref{3} we show that the addition of the emission component removes some of the required fine tuning in the constant density models, as the emission fills in part of the absorption trough. This effect becomes increasingly significant as the absorption trough gets deeper. This prevents a small change in the ionization state or the column density resulting in a large change in the absorption profile.

\begin{figure}
\centering
\begin{minipage}{85 mm}
\centering
\vbox{
\includegraphics[height=8 cm, angle=270]{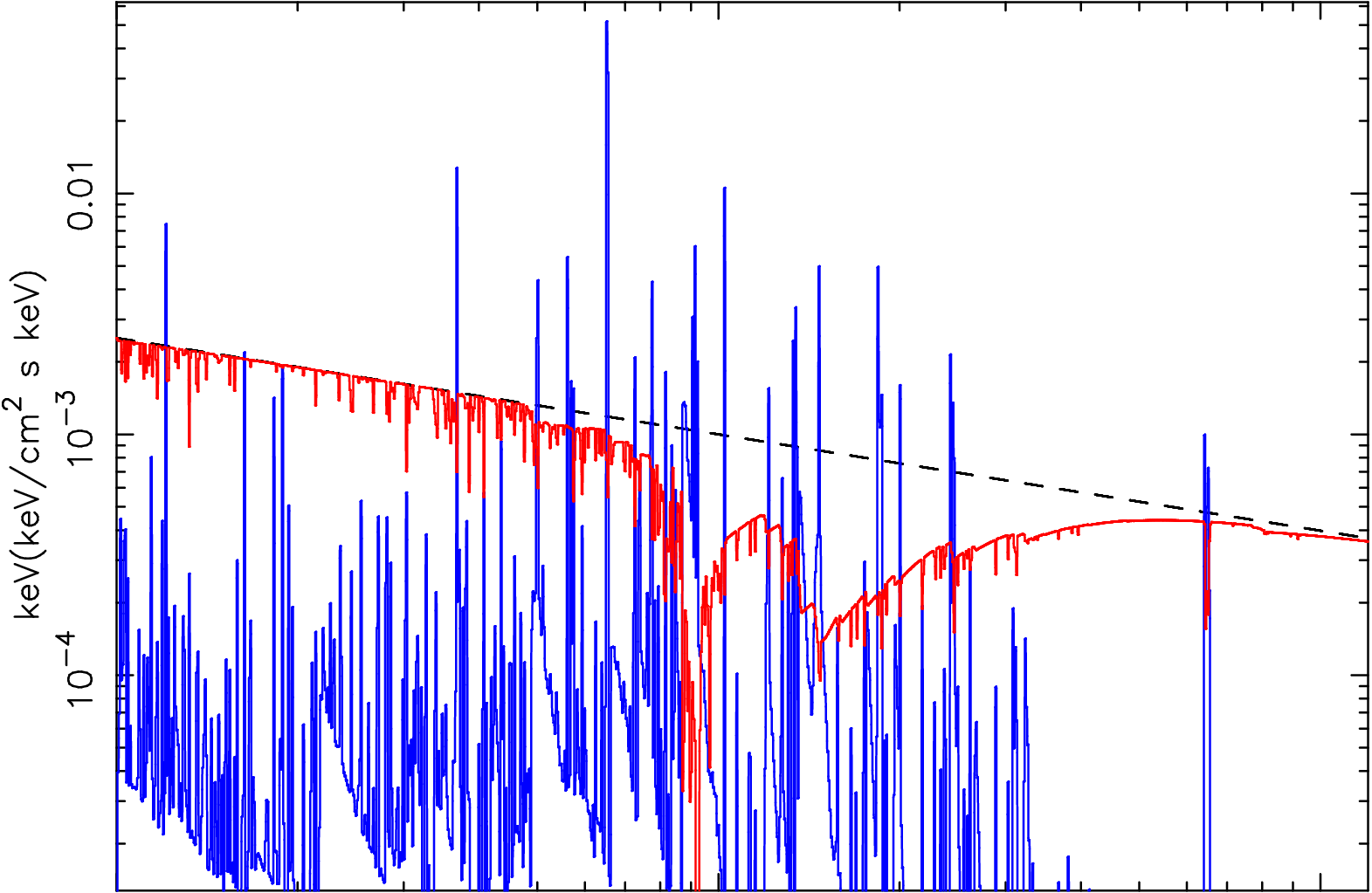}
\includegraphics[height=8 cm, angle=270]{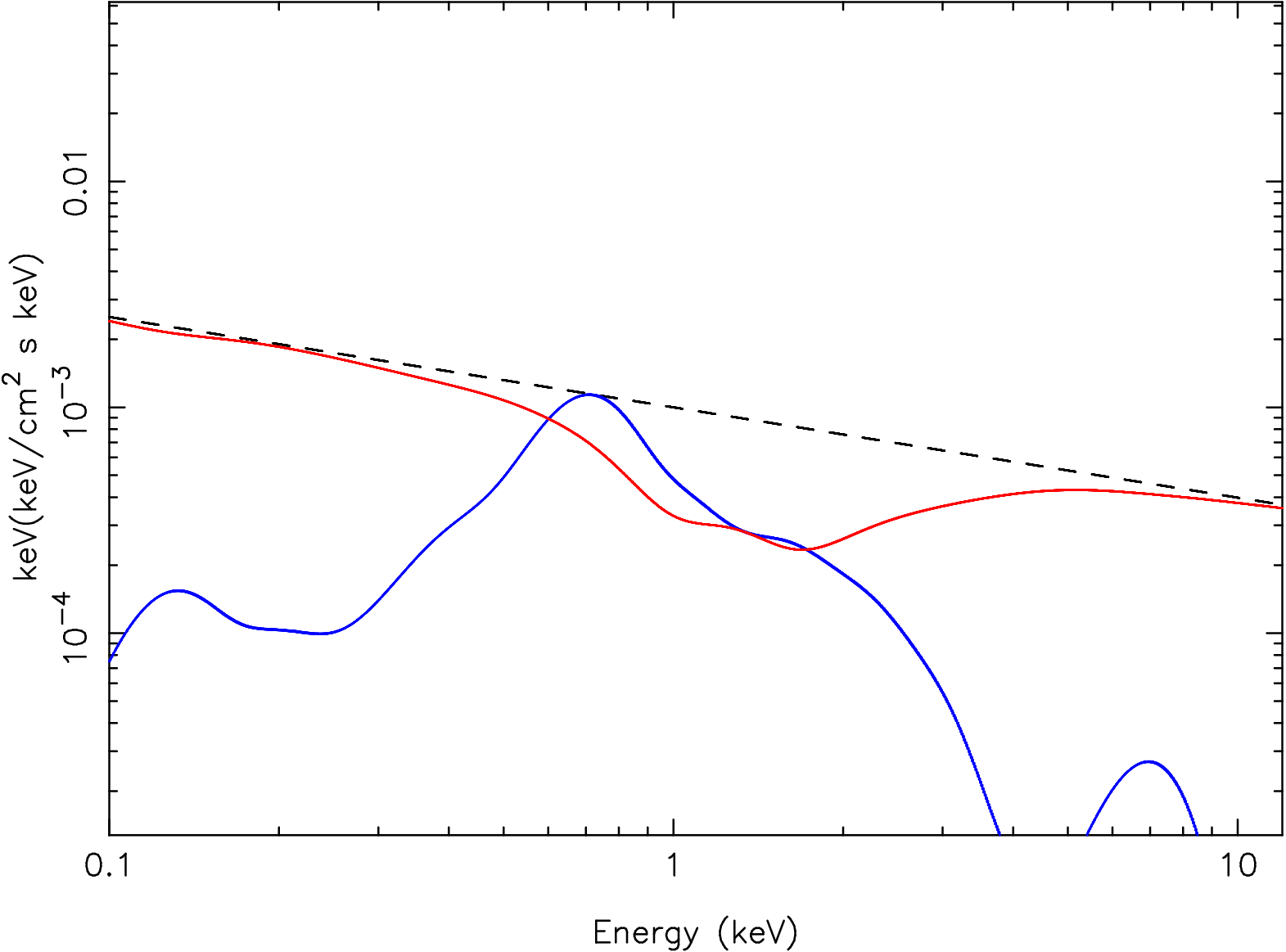}
}
\caption{{\it Upper Panel}: An example of the emission (blue) and absorption (red) X-ray spectra from gas photoionized by the X-ray continuum of a quasar. The intrinsic input spectrum (black, dotted) is a power-law with $\Gamma$=2.4 \& K$_{pl}$=10$^{-3}$ photons keV$^{-1}$ cm$^{-2}$ s$^{-1}$. The photoionized material has N$_{H}$=10$^{23}$ cm$^{-2}$ \& log$_{10}$($\xi$)=2.6, in addition to the details given in Section~\ref{2-1}. {\it Lower Panel}: The photoionized emission (blue) and absorption (red) spectra from the upper panel, smeared with a Gaussian velocity distribution with a typical {\it v/c}=0.2 The intrinsic input spectrum (black, dotted) is the same as in the upper Panel.}
\label{photoionmodel}
\end{minipage}
\end{figure}

The calculated spectra are dominated by O{\small VII}, O{\small VIII} and iron L-shell features seen in both emission (rest energies $\sim$0.55, 0.75 \& 0.7-1.5 keV respectively) and absorption (rest energies $\sim$0.74, 0.87 \& 0.73-1.55 keV respectively). In addition, the spectrum shows absorption associated with the iron M-shell unresolved transition array ($\sim$0.73-0.8 keV, Behar, Sako \& Kahn 2001). Figure~\ref{photoionmodel}, {\it Upper Panel}, shows an example emission and absorption spectrum from the grids of spectra calculated by XSTAR.

\subsection{Velocity structure}
\label{2-2}

Simulations of line-driven disk winds show a complex velocity structure (PK04), that is beyond the scope of this work to duplicate. The simulations demonstrate that the wind material typically accelerates in the radial direction, however there are several distinct velocity regions including a region in which the material in-falls, a low-density, rapidly accelerating region and a somewhat higher density, less-rapidly accelerating region. Further complicating matters, while the final velocity-smeared absorption profile depends only on the velocity distribution along specific line of sight, the final emission smearing profile depends on the global velocity distribution of all the emitting material. For this work, we make the simplifying assumption that the distribution of velocities present in the wind is a Gaussian, with a width that represents a typical velocity present in the wind (more precisely, 84\% of the ionized material in the wind has a velocity given by the value, $\sigma_{wind}$, or less). 

While this extremely simplistic assumption is unlikely to represent the correct profile in the case of absorption for a wind, as it results in as much red-shifted absorption as blue-shifted absorption, it may be more reasonable in the case of the total emission profile which will necessarily include emission from all parts of the wind. We note that changing the global velocity profile will result in significant differences in the absorption line profiles and, to a lesser extent, the emission line profiles.

Figure~\ref{photoionmodel}, {\it Lower Panel}, shows the effect of convolving the velocity distribution with the example XSTAR absorption and emission spectra and demonstrates that the intrinsically narrow features appear as continuum-like components if the velocity distribution is sufficiently broad. Interestingly, between 0.7 and 1.3 keV, the smeared emission spectrum rises somewhat more steeply than the curvature imprinted purely by absorption, suggesting that the emission will strongly affect the shape of the soft X-ray excess predicted by this model, provided the covering fraction is sufficiently large.

We note that this model is not the same as one in which the material has a large turbulent velocity. The amount of turbulent velocity present in absorbing material strongly affects the opacity of heavily saturated absorption lines (see {\it e.g.} Peterson 1997). A large turbulent velocity will broaden the saturated cores of these lines resulting in considerably more absorption than we might expect from this material. The impact of turbulence is discussed in more detail in Section~\ref{5}.

\section{A detailed examination of the wind properties}
\label{3}

Before applying the disk wind model it is instructive to understand how the model varies with the free parameters, in particular, how the column density (N$_{H}$), the ionization parameter ($\xi$) and the global covering fraction of the material (C$_{f}$ - see Section~\ref{3-3}) affect the emission spectrum and how significant this component is to the resulting total spectrum.

The total emitted luminosity is given by;

\begin{flushright}
\begin{minipage}{80mm}
\begin{equation}
{\hspace{5mm}}L_{emm}\,=\,C_{f}\,(L_{RRC}+L_{lines}+L_{ff})\,
\label{eq1}
\end{equation}
\end{minipage}
\end{flushright}

\hspace{-6mm}where $L_{lines}$ is the luminosity of the emission (bound-bound transitions), $L_{RRC}$ is the luminosity of the radiative recombination continua (free-bound transitions) and $L_{ff}$ is the luminosity of the free-free continuum (free-free transitions). All three emission mechanisms have the same dependence on the electron density of the emitting material ($n_{e}$) and the volume of the emitting material ($V$), which we can relate directly to the free parameters in out wind model. Specifically,

\begin{flushright}
\begin{minipage}{80mm}
\begin{displaymath}{\hspace{5mm}}L_{RRC}\,=\,n_{e}^{2}\,V\,F_{RRC}(\xi,h\nu,\alpha_{RRC})\end{displaymath}
\begin{displaymath}{\hspace{5mm}}L_{lines}\,=\,n_{e}^{2}\,V\,F_{lines}(\xi,h\nu,\alpha_{lines}) \end{displaymath}
\begin{equation}{\hspace{5mm}}L_{ff}\,=\,n_{e}^{2}\,V\,F_{ff}(\xi,h\nu,\alpha_{ff})
\label{eq2}
\end{equation}
\end{minipage}
\end{flushright}

\hspace{-6mm}where $F_{RRC}$, $F_{lines}$ \& $F_{ff}$ are functions of the ionization state of the material, $\xi$, the energy released by the interactions involved, $h\nu$, and a coefficient appropriate to each type of interaction, $\alpha_{...}$ (see Appendix~\ref{app-1}).

For the XSTAR models used here, the volume of the emitting material is $V$=$4\pi r^{2}\Delta R$, where $r$ is the radius from the central source to the inner edge of the wind material and $\Delta R$ is the radial depth of the shell. In addition, $\Delta R$ can be defined in terms of the density and the column density of the material such that $\Delta R=N_{H}/n$. Combining these with Equations~\ref{eq1} \&~\ref{eq2} we have,

\begin{flushright}
\begin{minipage}{80mm}
\begin{equation}
{\hspace{5mm}}L_{emm}\,=\,C_{f}\,n_{e}\,4\pi r^{2}\,N_{H}\,(F_{RRC}+F_{lines}+F_{ff})
\label{eq3}
\end{equation}
\end{minipage}
\end{flushright}

Furthermore, the ionization state of the material is given by $\xi=L_{int}/n_e r^{2}$, where $L_{int}$ is the intrinsic luminosity of the ionizing continuum. Combining this with Eq~\ref{eq3} gives;

\begin{flushright}
\begin{minipage}{80mm}
\begin{equation}
{\hspace{5mm}}{L_{emm} \over L_{int}}\,=\,{C_{f}\,N_{H} \over \xi}\,4\pi(F_{RRC}+F_{lines}+F_{ff})
\label{eq4}
\end{equation}
\end{minipage}
\end{flushright}

\hspace{-6mm}showing that, to zeroth order, the luminosity in the emission, as a fraction of the intrinsic ionizing luminosity, is proportional to the column density and the covering fraction and inversely proportional to the ionization state. We emphasise that the dependence on ionization state {\it is} more complex than the simple inverse relation implied by the initial part of Eq~\ref{eq4}, in that the possible atomic transitions are different for different ionization states of the gas, giving very different values for each of the $F_{n}$-functions as we vary $\xi$. Building on the simple relationships in Eq~\ref{eq4}, we examine the impact of, and the interplay between, the emission and absorption on the shape of the total X-ray spectrum, as we vary the N$_{H}$, $\xi$ and $C_{f}$.

\subsection{Column density}
\label{3-1}

\begin{figure}
\centering
\begin{minipage}{85 mm}
\centering
\vbox{
\includegraphics[height=8 cm, angle=270]{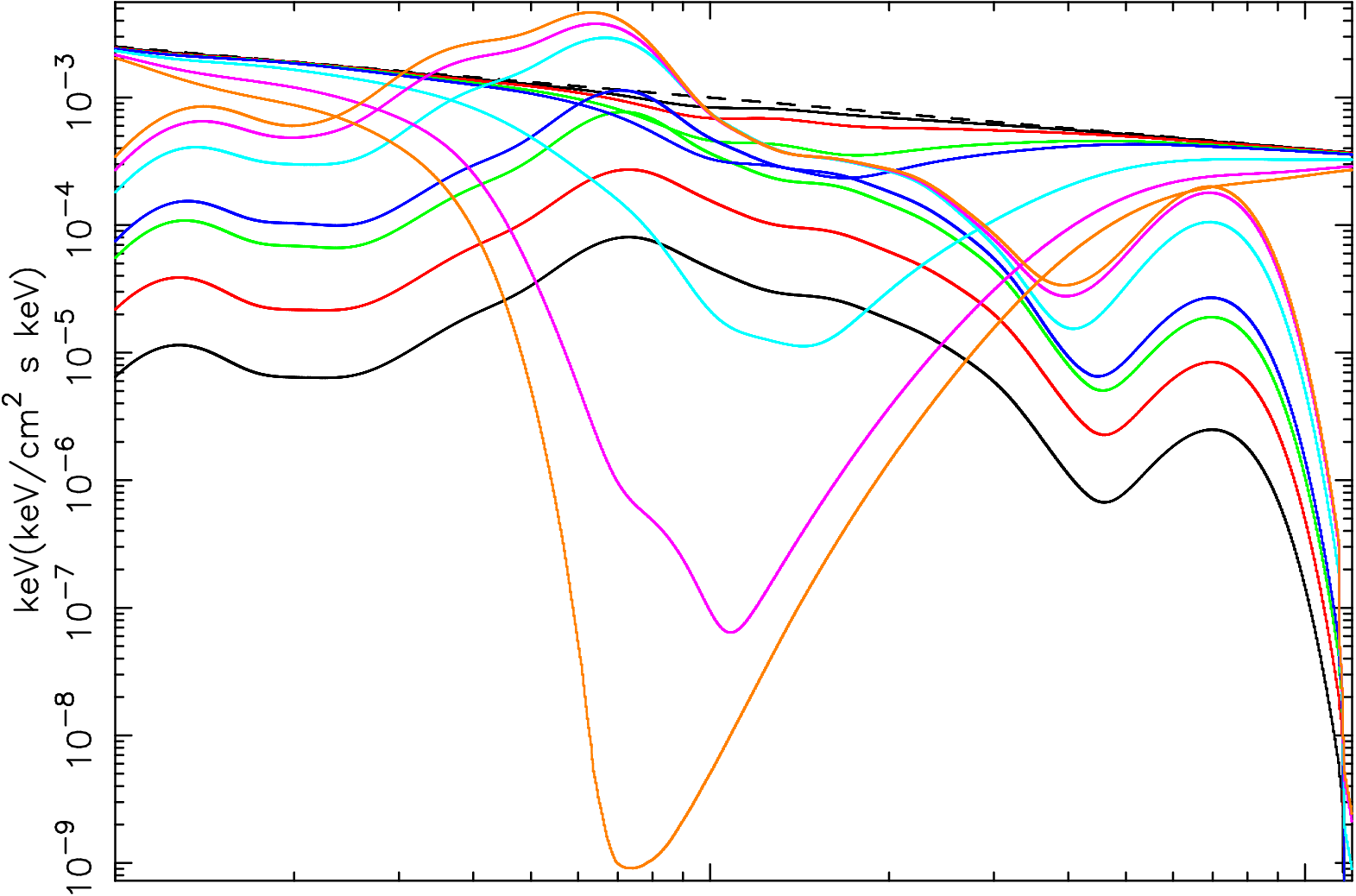}
\includegraphics[height=8 cm, angle=270]{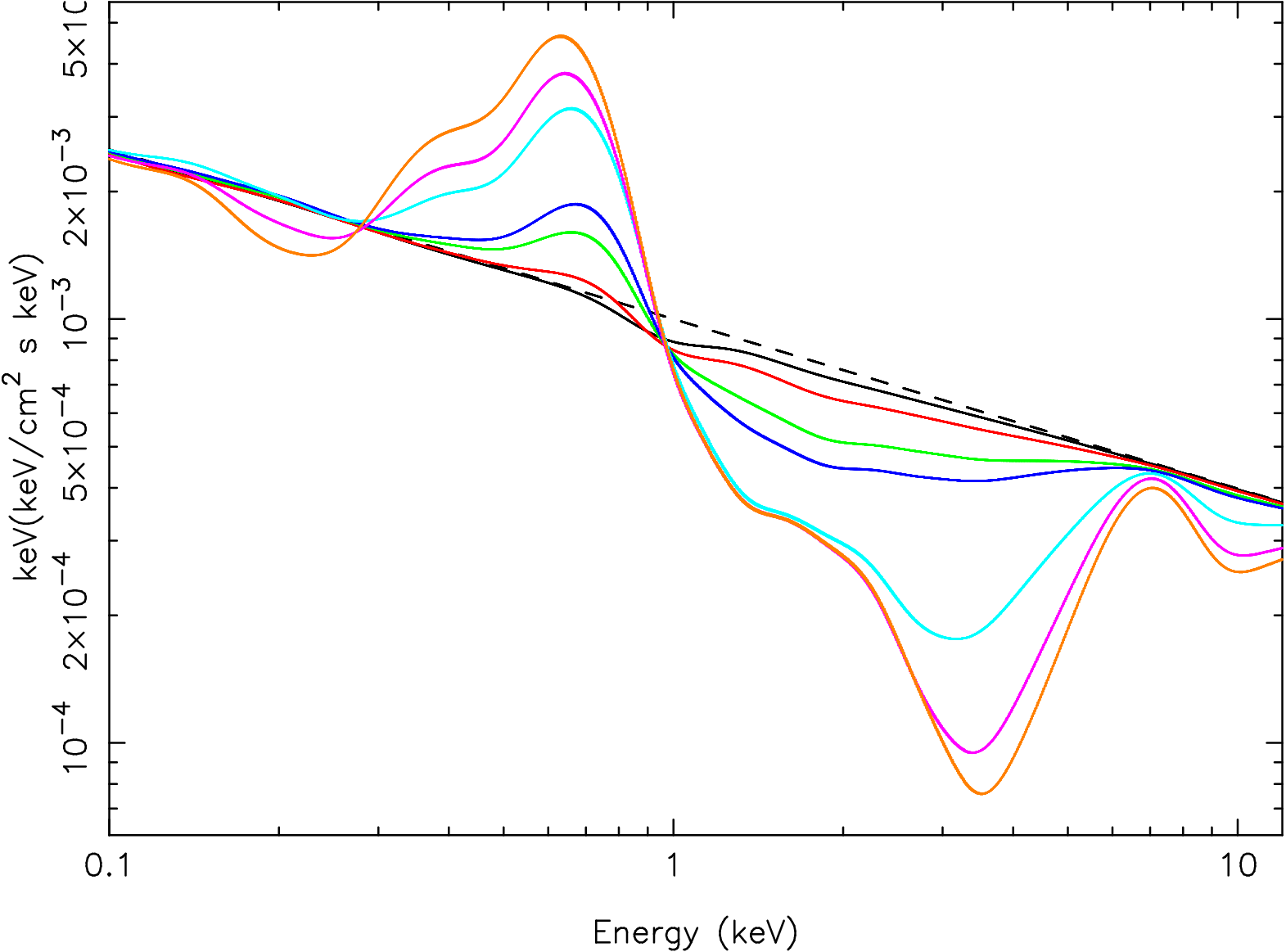}
}
\caption{The effects of varying N$_{H}$ on the smeared wind model. {\it Upper Panel}: The dotted line represents the intrinsic input continuum (power-law, $\Gamma$=2.4, K$_{pl}$=10$^{-3}$ photons keV$^{-1}$ cm$^{-2}$ s$^{-1}$). The solid lines show the smeared absorption and emission spectra (log$_{10}$($\xi$)=2.6, $\sigma$$_{wind}$=0.2, C$_{f}$=1.0) for column densities of 1, 3.3 \& 6.6 $\times$10$^{22}$ (black, red \& green), 1, 3.3 \& 6.6 $\times$10$^{23}$ (dark blue, light blue \& purple) \& 1$\times$10$^{24}$ (brown) cm$^{-2}$. {\it Lower Panel}: The solid lines show the `total' spectrum (absorption$\times$intrinsic spectrum + emission) for the same set of parameters.}
\label{Nhcomp}
\end{minipage}
\end{figure}

Figure~\ref{Nhcomp} shows the spectral shape changes associated with changing the column density. The upper panel shows the effect of changing N$_{H}$ on the absorption and emission spectra, while the lower panel shows the effect on the total spectrum. Below N$_{H}$$\sim$5$\times$10$^{22}$ cm$^{-2}$, even with a covering fraction of one, the smeared emission is not a significant correction to the shape of the absorbed spectrum. Above this however, the emission becomes increasingly important, eventually becoming the dominant spectral component between 0.2-3 keV at column densities of $\sim$10$^{24}$ cm$^{-2}$.

The absorption increases dramatically with increasing column, as expected, but there is also a more subtle effect, namely that the trough of the absorption component shifts to lower energies with increasing column, indicating the presence of lower ionization material. This is a natural consequence of the increase in absorption in the front layers of the cloud reducing the flux transmitted deeper within the cloud (the reduction in flux due to the increasing distance is negligible since the cloud has $\Delta$R$<<$R). Thus increasing the column leads to more shielding and so to progressively lower ionization material at the back of the cloud.

The emission also follows this pattern, with the emitted flux increasing strongly with column, but also showing a subtle shift to slightly lower ionization states, as revealed by the slightly lower peak energy (due to the strong O{\small VII}/O{\small VIII} lines) Interestingly, the emission is always characterised by a sharp rise below $\sim$1 keV, precisely the point where we observe the sharp rise of the soft X-ray excess. This suggests that the emission component may, in fact, assist us in accounting for the soft X-ray excess! Intruigingly, similar models of emission (though not absorption) from photoionised material were used to fit the soft excess seen in the X-ray spectra of many accreting pulsars (Hickox, Narayan \& Kallman 2004). 

The co-added emission plus absorption spectra shown in the lower panel of Figure~\ref{Nhcomp} support this, showing that the infill from the emission offsets the effect of lower ionization material at high columns, leading to an apparent rise in the spectrum which is always close to $\sim$1 keV. Significantly, however, the peak in the emission is quite obvious in these model spectra, leading to a pronounced dip below $\sim$0.6 keV. This should be an observable consequence of a large contribution from the emission of the material. While such a dip is not commonly reported in the shape of the soft X-ray excess in AGN, this could be masked by the presence of either a Compton reflection continuum, {\em narrow} ionized absorption intrinsic to the AGN, neutral absorption intrinsic to the AGN and/or galactic absorption (typically 10$^{20-21}$ cm$^{-2}$).

\subsection{Ionization state}
\label{3-2}

\begin{figure}
\centering
\begin{minipage}{85 mm}
\centering
\vbox{
\includegraphics[height=8 cm, angle=270]{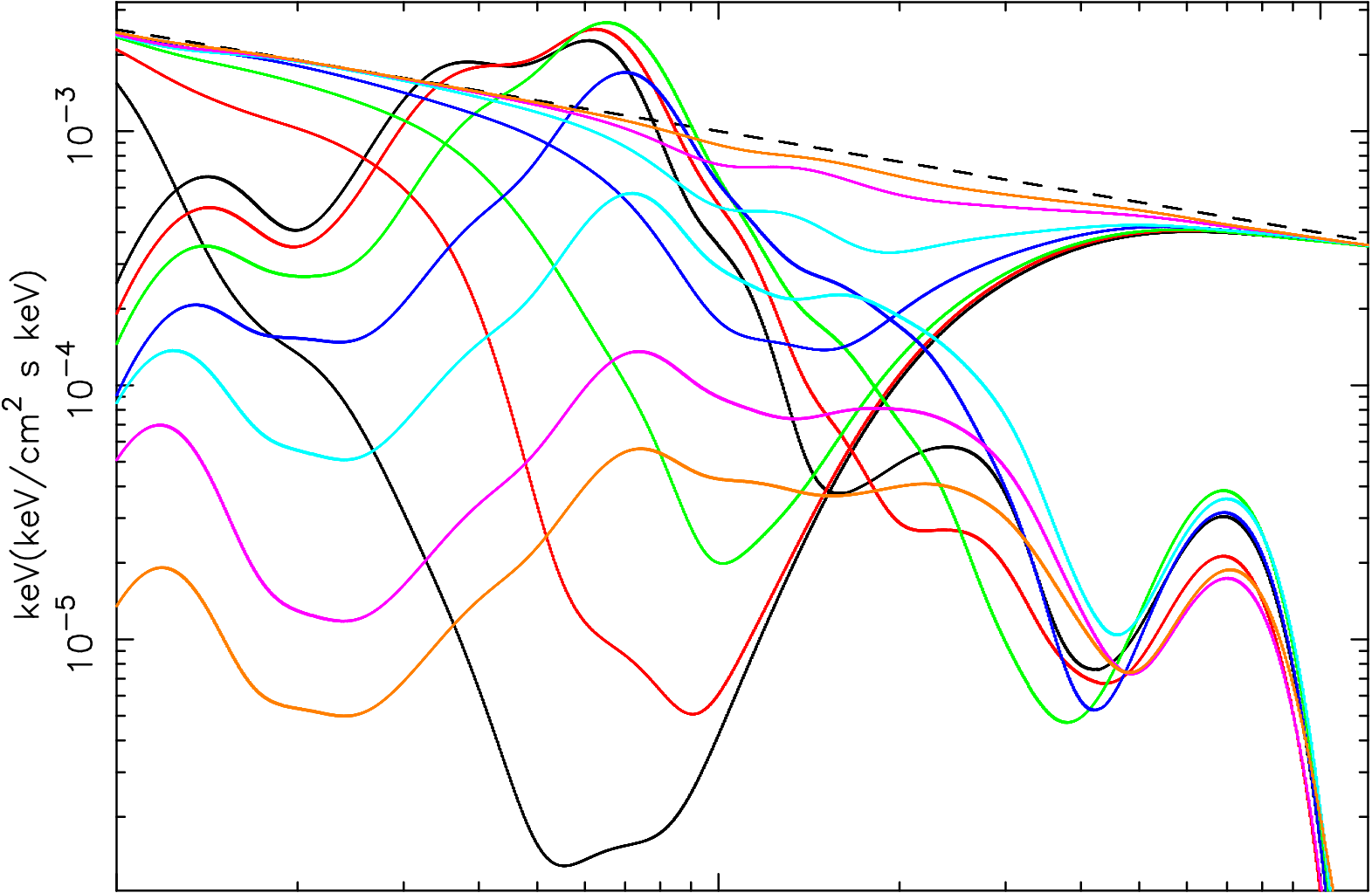}
\includegraphics[height=8 cm, angle=270]{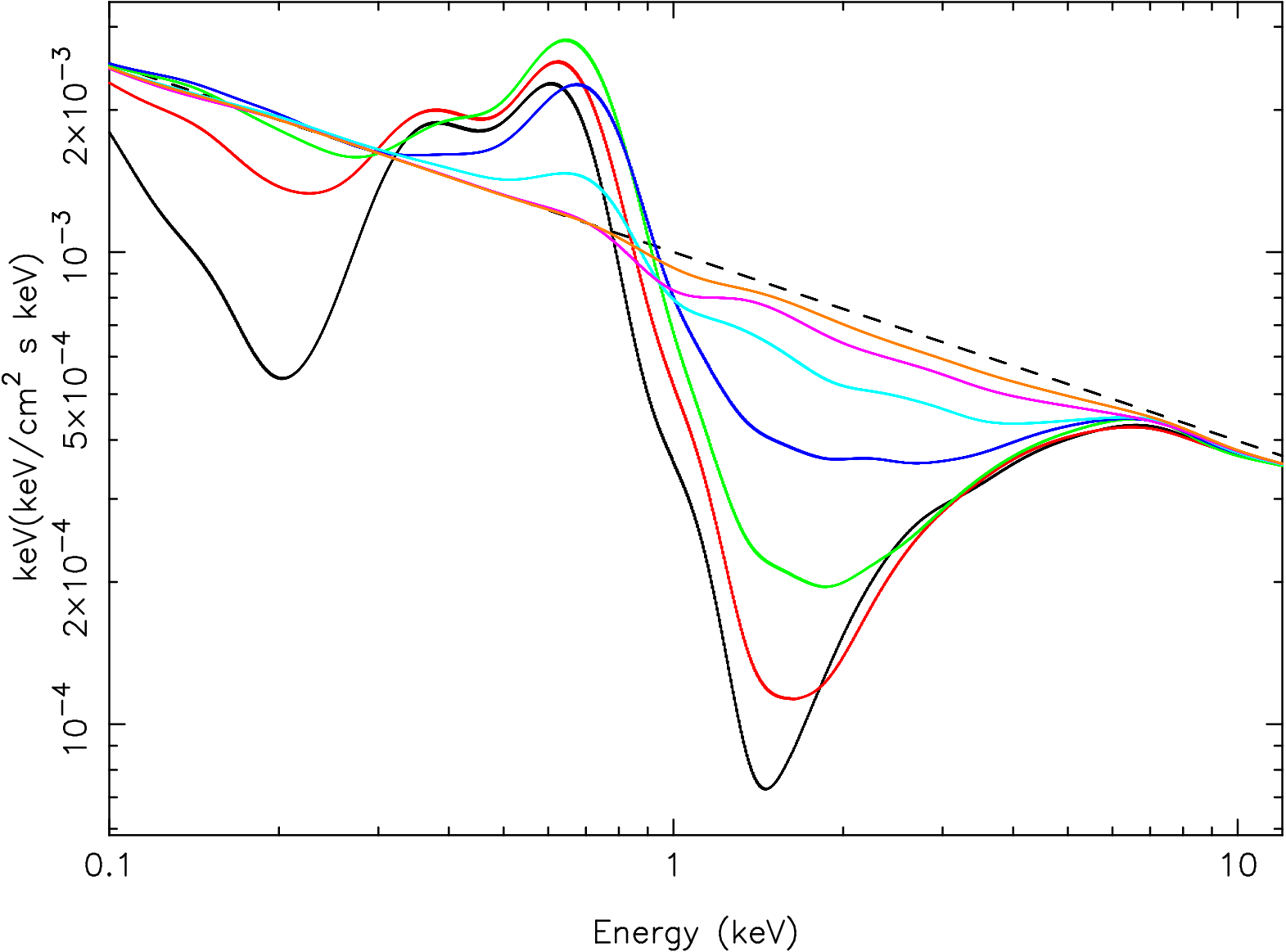}
}
\caption{The effects of varying $\xi$ on the smeared wind model. {\it Upper Panel}: The dotted line represents the same intrinsic input continuum detailed in Fig~\ref{Nhcomp}. The solid lines show the smeared absorption and emission spectra (N$_{H}$=10$^{23}$ cm$^{-2}$, $\sigma$$_{wind}$=0.2, C$_{f}$=1.0) with ionization states of log$_{10}\xi$=1.5, 1.83, 2.16, 2.5, 2.83, 3.16 \& 3.5 (black, red, green, dark blue, light blue, purple \& brown). {\it Lower Panel}: The solid lines show the total spectrum for the same set of parameters.}
\label{xicomp}
\end{minipage}
\end{figure}

Figure~\ref{xicomp} shows the spectral shape changes associated with changing the ionization state. The upper panel shows the effect of changing $\xi$ on the absorption and emission spectra, while the lower panel shows the effect on the total spectrum. The effects of increasing $\xi$ are broadly opposite to the effects of decreasing the column density (as we would expect from equation~\ref{4}), in that the absorption trough becomes less pronounced and shifts to higher energies. Physically, the opacity of ionized material falls with increasing ionization state and the material becomes dominated by higher ionization species of each ion, giving rise to less pronounced, higher energy, absorption troughs at higher ionization states.

While increasing the column led to some lower ionization material present at the back of the cloud, this is a very small effect compared to lowering the ionization parameter at the front of the cloud. Fig.~\ref{xicomp} shows that the low energy absorption increases substantially below log$_{10}\xi\sim$2.1, mainly due to the presence of L shell opacity from Oxygen VI and Nitrogen V as well as Fe M shell features.

The effect of varying $\xi$ on the emission is considerably more `messy' than varying $N_{H}$. Specifically, the contribution of N {\small VII}, O{\small VII} \& O{\small VIII}, relative to highly ionized species of sodium, magnesium and silicon, drops dramatically above ionization states of log$_{10}$($\xi$)$\sim$2.8. This removes the peak in the emission at $\sim$0.6 keV due to O{\small VII} line emission, so the spectrum is much flatter at these ionization states.

The total spectrum (seen in the lower panel of Fig.~\ref{xicomp}) shows again that including the emission counteracts some of the sensitivity to the parameters predicted from absorption alone. There is a soft excess which rises at $\sim$1 keV for log$_{10}$($\xi$)=1.8-3. Again, a key characteristic of the emission from ionization states below log$_{10}$($\xi$)$\sim$2.6 is that it peaks at $\sim$0.6 keV, and these models predict a dip at lower energies if the emission is important.

Including the self-consistent emission from the partially ionized material results in a stable soft excess for a wider range of log$_{10}$($\xi$) than expected from pure absorption alone. We note that the slope of the smeared emission spectrum between 0.7-2 keV is relatively sensitive to changes in ionization state which may explain the distribution of soft excess shapes observed in AGN (Crummy \etal 2005).

\subsection{Covering fraction}
\label{3-3}

As discussed in Section~\ref{2-1}, the properties of the absorption seen along the line-of-sight (l.o.s) are, to first order, independent of the covering fraction of the material, however the relative strength of the emission is sensitive to the covering fraction. In fact, the absorption spectrum {\it does} depend weakly on the {\it local} covering fraction ({\it i.e.}, whether the material in the immediate environment of the l.o.s is clumpy or uniformly smooth) due to some subtleties in the way XSTAR propagates radiation from one layer in the material to the next (Kallman {\it priv. comm.}). By contrast, the emission is sensitive to both the global covering fraction ({\it i.e.}, whether the global distribution of the material is in a uniform shell, a non-uniform shell, a conical geometry, etc) and the local covering fraction. The covering fraction variable available in XSTAR ({\it cfrac}) strictly refers to the local covering fraction. The global covering fraction is, by default, unity for a single XSTAR run, since XSTAR simulates a spherical shell of material, however the grid of calculated emission and absorption spectra are stored as separate table model and are treated individually by XSPEC. In this case, the global covering fraction is folded into the normalization of the emission additive model.

\subsubsection{The normalization of the emission}
\label{3-3-1}

It is instructive to understand what the normalization of the emission component actually means before we use it in XSPEC. The normalization of the emission ($K_{emm}$) is given by;

\begin{flushright}
\begin{minipage}{80mm}
\begin{equation} {\hspace{5mm}}K_{emm}\,=\,{C_{f}\,L_{int} \over 10^{38}}{1 \over D^{2}_{kpc}}
\label{eq5}
\end{equation}
\end{minipage}
\end{flushright}

\hspace{-8mm} where $C_{f}$ is the global covering fraction and $D_{kpc}$ is the distance to the source in kpc (this expression is derived in detail in the XSTAR 2.1kn3 Manual, $\S$ 7.2.3.\footnote{{\small http://heasarc.gsfc.nasa.gov/docs/software/xstar/docs/html/ node8.html}}). For a power-law continuum of the form N(E)=$K_{pl}$$(E/1{\rm keV})^{-\Gamma}$, where $E$ is energy in keV and $K_{pl}$ is the power-law normalization at 1keV in photons keV$^{-1}$ cm$^{-2}$ s$^{-1}$, $L_{int}$ is given by;

\begin{flushright}
\begin{minipage}{80mm}
\begin{equation} {\hspace{5mm}}L_{int}\,=\,4\pi X\,D^{2}_{kpc}\,K_{pl}\,{(E^{-\Gamma+2}_{2}\,-\,E^{-\Gamma+2}_{1}) \over (-\Gamma+2)}
\label{eq6}
\end{equation}
\end{minipage}
\end{flushright}

\hspace{-8mm} where the constant $X$ contains the conversion from cm$^{2}$ to kpc$^{2}$ and from keV to erg (=1.53$\times$10$^{34}$), $E_{2}$ \& $E_{1}$ are the upper and lower limits of the energy range respectively (1-1000 Rydbergs in XSTAR). Combining Eq~\ref{eq5} \&~\ref{eq6}, the distance dependence of the emission normalization can be removed and we obtain an equation that depends solely on the details of the input power-law continuum (specifically its normalization and its slope) and the global covering fraction of the ionized material.

\begin{flushright}
\begin{minipage}{80mm}
\begin{equation} {\hspace{5mm}}K_{lines}\,=\, {4\pi X \over 10^{38}}\,C_{f}\,K_{pl}\,{(E^{-\Gamma+2}_{2}\,-\,E^{-\Gamma+2}_{1}) \over (-\Gamma+2)}
\label{eq7}
\end{equation}
\end{minipage}
\end{flushright}

Clearly, allowing the normalization of the emission to be independent of the normalization of the underlying continuum in any actual fit to real data, gives us a measure of the global covering fraction of the ionized material. In the remainder of this work, we assume that the ionized material is uniformly smooth on all scales and interpret the global covering fraction as an indicator of the large scale geometry of the wind material.

\subsubsection{The effect of changing $C_{f}$}
\label{3-3-2}

\begin{figure}
\centering
\begin{minipage}{85 mm}
\centering
\vbox{
\includegraphics[height=8 cm, angle=270]{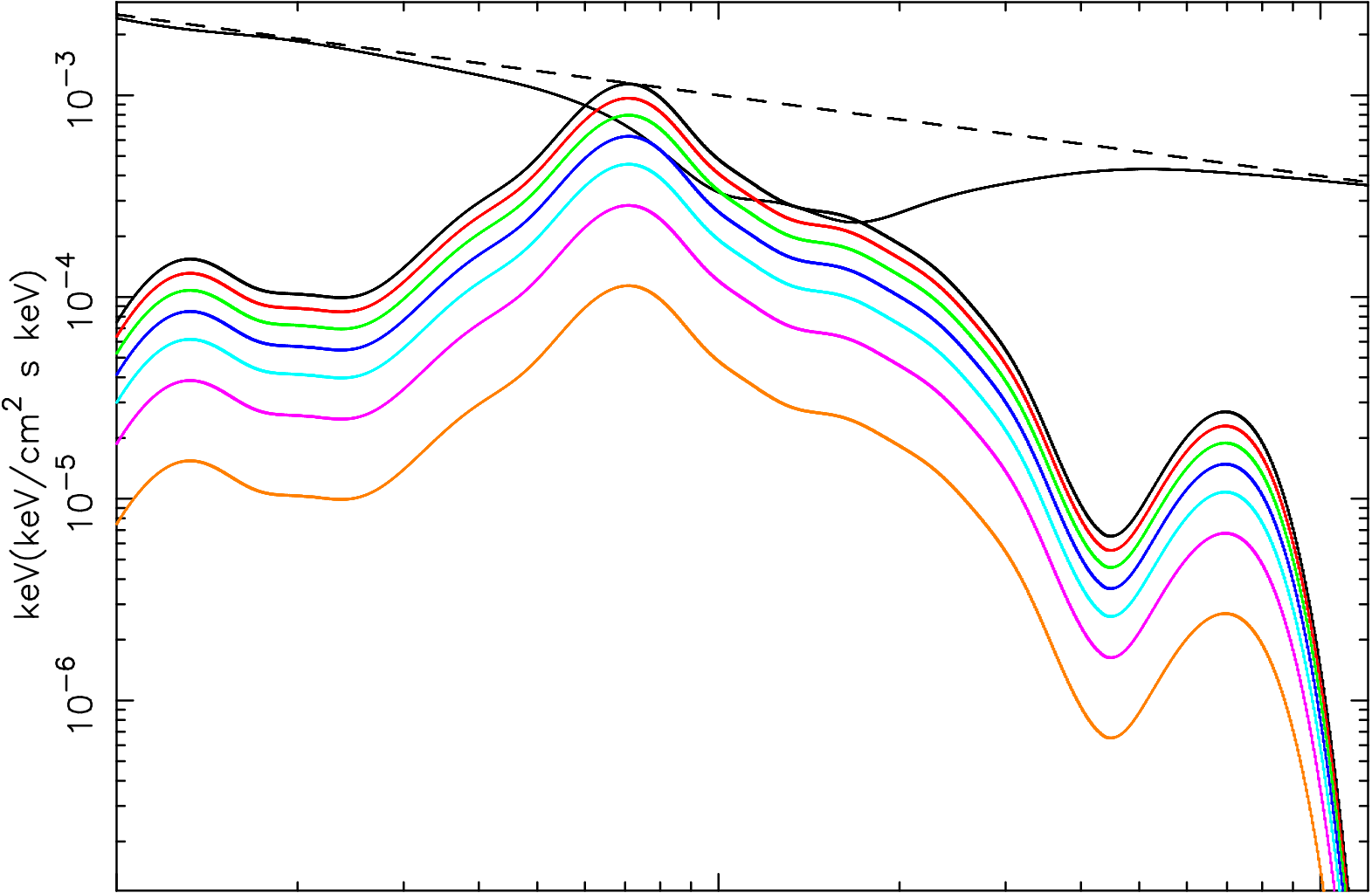}
\includegraphics[height=8 cm, angle=270]{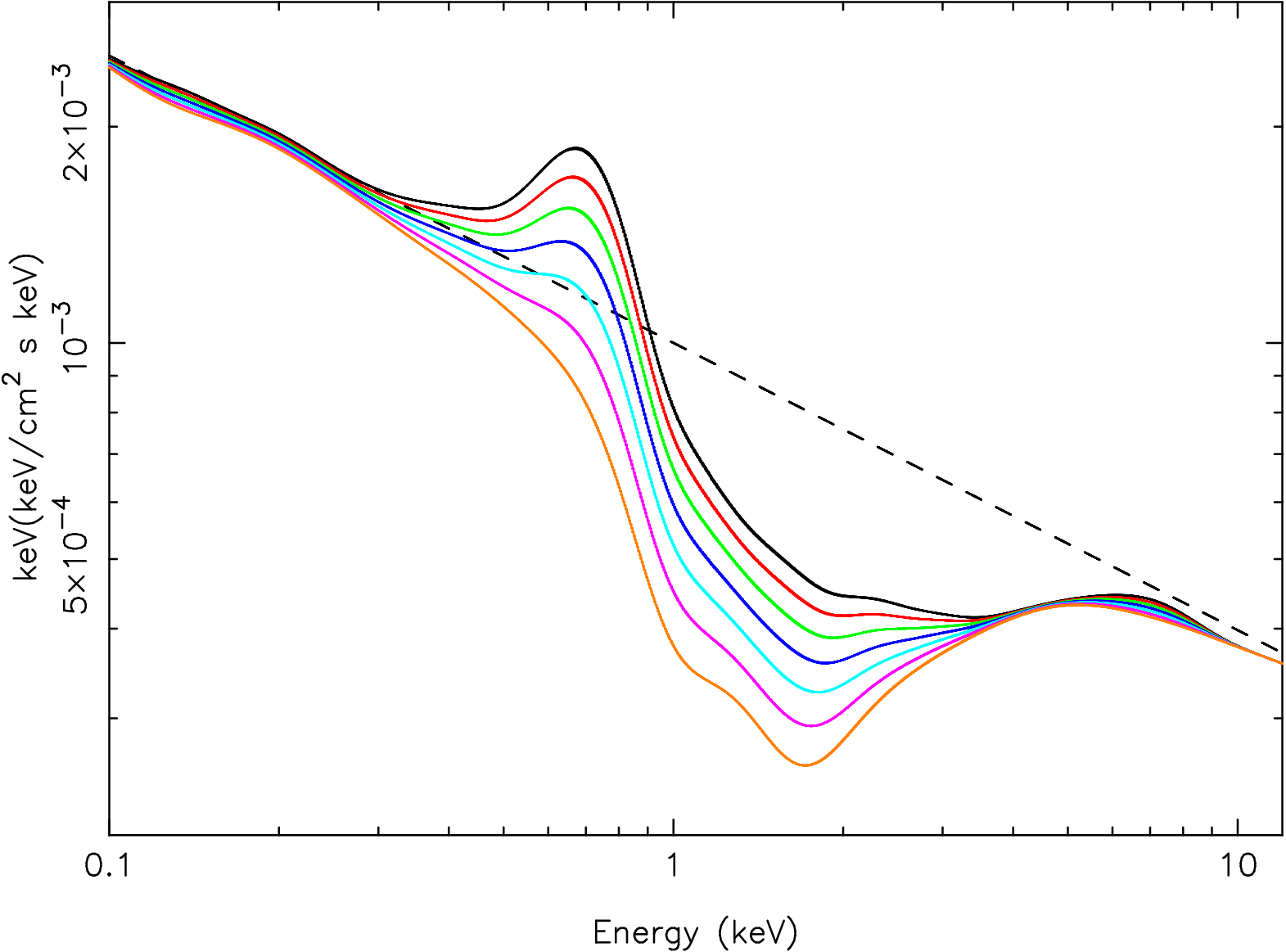}
}
\caption{The effects of varying C$_{f}$ on the smeared wind model. {\it Upper Panel}: The dotted line represents the same intrinsic input continuum detailed in Fig~\ref{Nhcomp}. The solid lines show the smeared absorption and emission spectra (N$_{H}$=10$^{23}$ cm$^{-2}$, log$_{10}$($\xi$)=2.6, $\sigma$$_{v}$=0.2) with covering fractions of 1.0, 0.85, 0.7, 0.55, 0.4. 0.25 \& 0.1 (black, red, green, dark blue, light blue, purple \& brown). {\it Lower Panel}: The solid lines show the total spectrum for the same set of parameters.}
\label{Cfcomp}
\end{minipage}
\end{figure}

Figure~\ref{Cfcomp} shows the spectral shape changes associated with changing the covering fraction of the ionized material. The upper panel shows the effect of changing C$_{f}$ on the emission component (the absorption component being independent of the global covering fraction), while the lower panel shows the effect on the total spectrum. The emission has a considerable effect on the spectrum even at relatively low covering fractions, particularly on the depth of the total observed absorption trough, however it is only at covering fractions of $\ga$50\% that the emission begins to dominate the spectrum. The characteristic `double-humped' absorption profile becomes considerably less pronounced as the emission begins to fill in the absorption at low energies. We note, again, that regardless of the relative strength of the emission the spectrum in the 0.7-1.5 keV range shows a steep rise reminiscent of the soft excess. Again, when the emission becomes very strong the spectrum shows an increasingly pronounced dip in the total spectrum below $\sim$0.6 keV, similar to the dips discussed previously for $N_{H}$ \& $\xi$.

\section{PG1211+143; A case study}
\label{4}

The objects which will provide us with the tightest constraints on the wind parameters are the bright quasars with particularly large column densities. On this basis we choose PG1211+143 on which to test this model. 

PG1211+143 is a bright (V$_{mag}$=14.38), nearby (z=0.089) quasar that has been identified as a Narrow Line Seyfert 1. The X-ray spectrum of PG1211+143 is known to be complex. \xte~observations suggest a significant Compton reflection component (Janiuk \etal 2001). \asca~observations revealed a strong soft X-ray excess and evidence that the iron K$\alpha$ emission line was broad (Reeves \etal 1997). \xmm~observations revealed the presence of several complex absorption systems that can either be interpreted as high velocity ($\sim$24,000 km s$^{-1}$), highly ionized absorption systems (Pounds \etal 2003), or lower velocity ($\sim$3,000-6,000 km s$^{-1}$) systems with a somewhat lower ionization state (Kaspi \etal 2005). 

Most importantly, PG1211+143 has one of the strongest soft X-ray excesses observed, with the X-ray flux below 0.6 keV being up four times greater than the flux predicted by continuum fits to the 1-10 keV spectrum (Pounds \etal 2003). Such a large soft excess makes it the ideal test of models for the origin of this component and as such it was used by GD04 as a test of their absorption-only model. Here we extend their fits to include the smeared emission as well as absorption, testing the ability of the full model to fit the soft excess seen in this object.

\subsection{Data Reduction}
\label{4-1}

\xmm~observed PG1211+143 in 2001 for 55ks (Obs ID: 0112610101). During the observation the EPIC PN was operated in Large Window Mode with the medium filter. The EPIC PN data were reduced and processed with the Science Analysis System (SAS v6.5.0) using the standard processing chain and the most up to data calibration data available as of November 2005. Periods of high background were excluded from the analysis and the spectrum was extracted from a 45\arcs extraction region resulting in a net exposure time of $\sim$49.5 ks. The spectrum was background subtracted and binned to have a minimum of 20 counts per bin to allow the use of chi-squared fitting statistics.

\subsection{Fitting the X-ray spectrum}
\label{4-2}

We begin by defining a model which accounts for all the complexity known to be present in the 0.3-12.0 keV spectrum of PG1211+143. We use a power-law continuum modified by Compton reflection, two un-smeared, partially ionized absorption systems, together with Galactic absorption (2.85$\times$10$^{20}$ cm$^{-2}$ -
Murphy \etal 1996). 

It is unclear from either Pounds \etal (2003) or Kaspi \etal (2006) which of their interpretations for the nature absorption complexities best represents the data. However, given that our aim is to test whether the accretion disk wind model can represent the broad shape of the X-ray spectrum, the details of the complex absorption systems are not particularly relevant to the work presented here, provided that they are sufficiently well modelled that they do not damage the quality of the overall fit. To this end, we adopt the interpretation of Pounds \etal on the basis of simplicity (since this method does not require ion by ion fitting). We construct two sets of XSTAR table models covering the parameter ranges detailed in Pounds \etal (N$_{H,1}$=5$\times$10$^{23}$ cm$^{-2}$, $\xi_{1}\sim$3.4, N$_{H,2}$=6$\times$10$^{21}$ cm$^{-2}$, $\xi_{2}\sim$1.7). We include the emission associated with each of these narrow absorption systems and allow the normalization of each of the emission components (equivalent to the covering factor of the material) to be a free parameter in the model.

\begin{figure}
\centering
\begin{minipage}{85 mm}
\centering
\vbox{
\includegraphics[height=8 cm, angle=270]{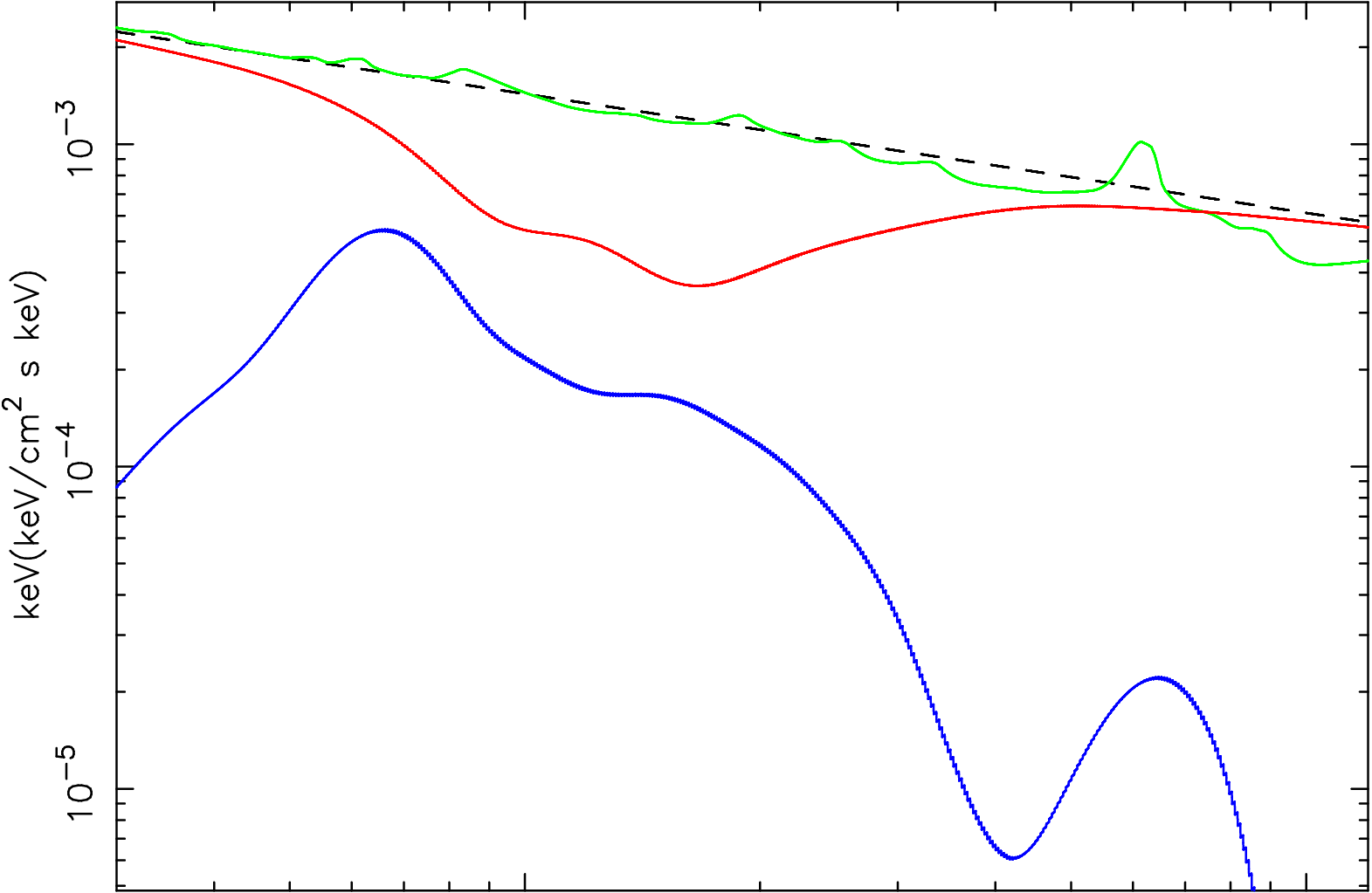}
\includegraphics[height=8 cm, angle=270]{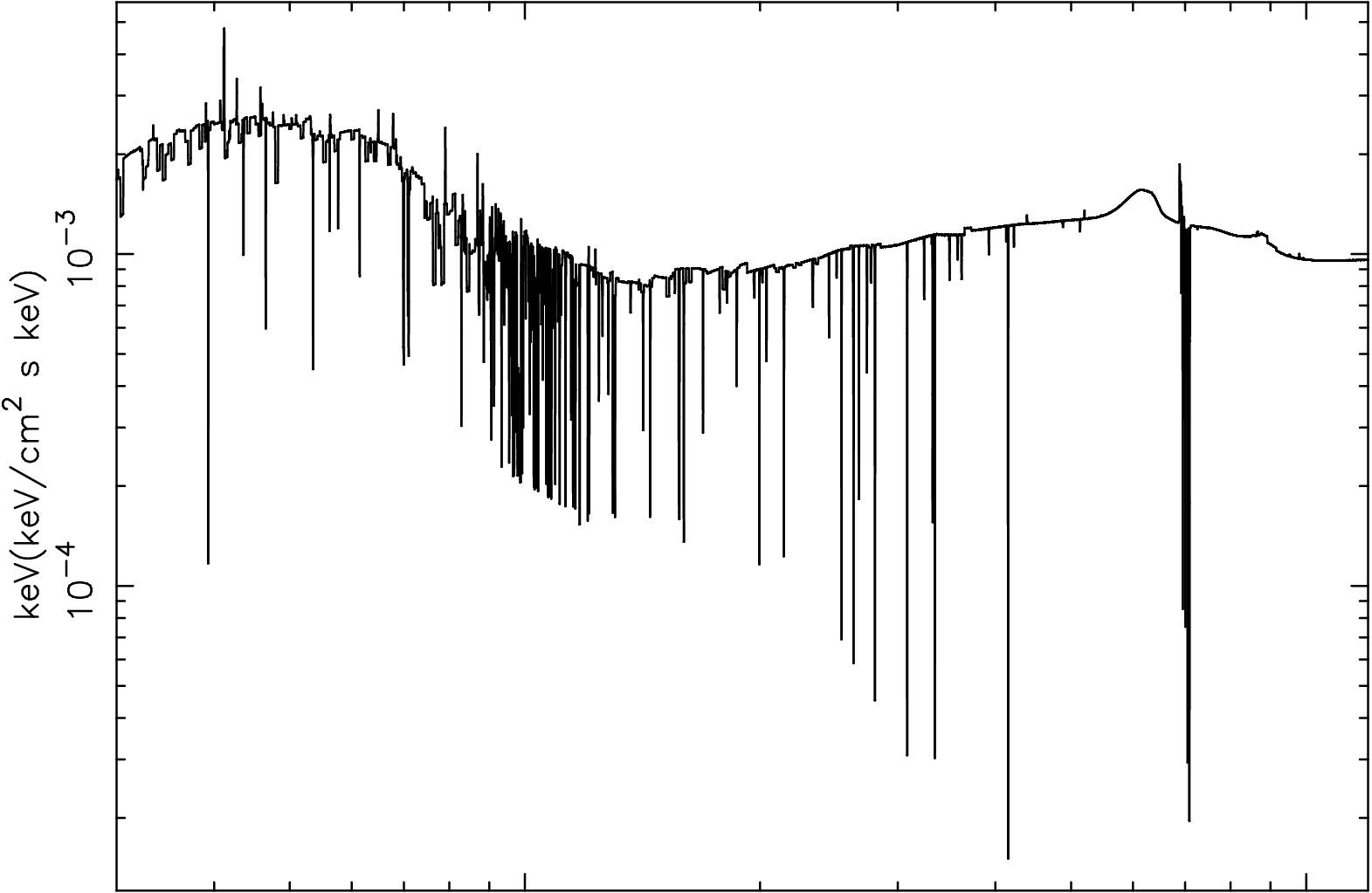}
\includegraphics[height=8 cm, angle=270]{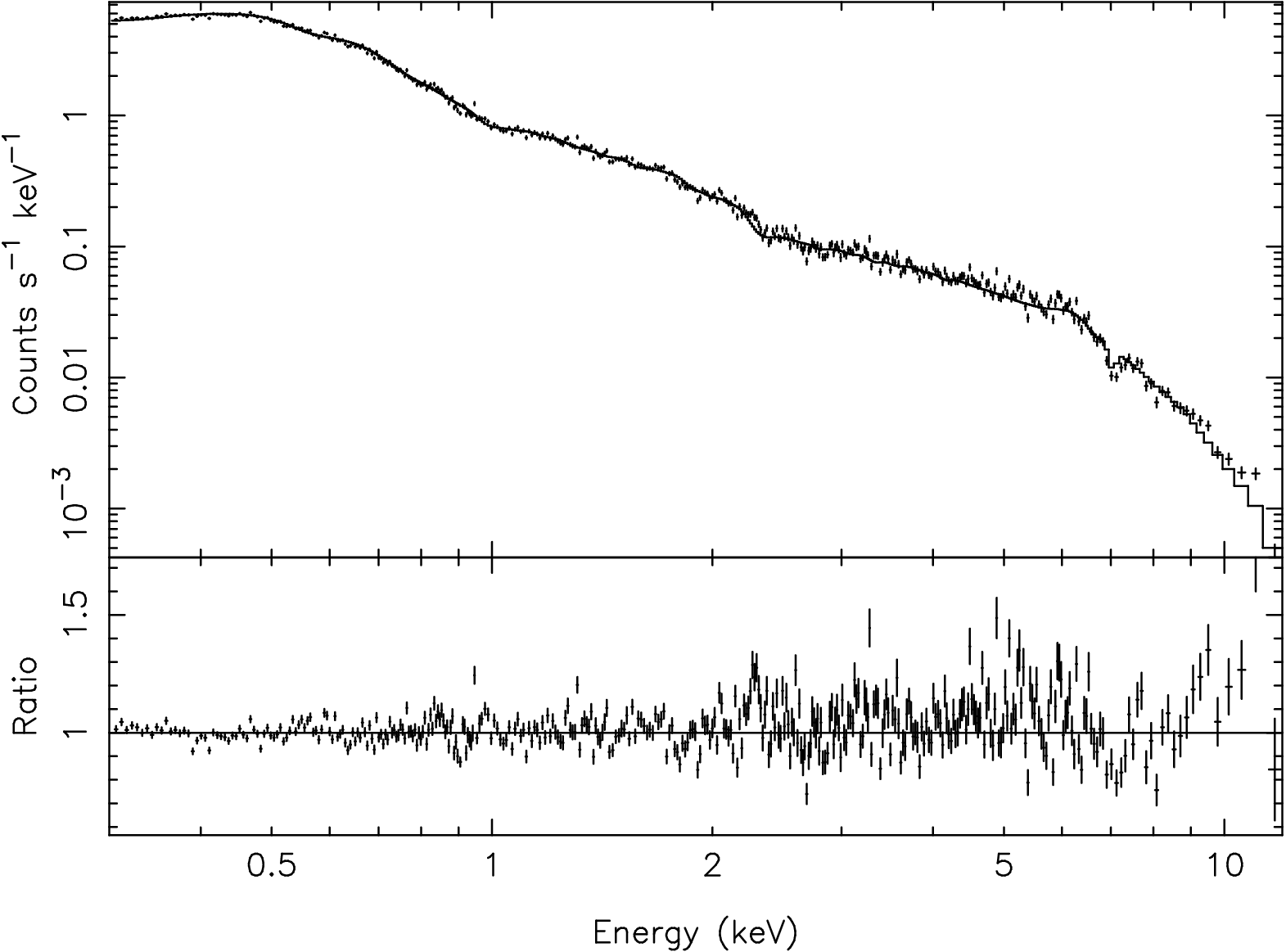}
}
\caption{{\it Top Panel} The best-fit model continuum components, namely the underlying power-law continuum (dashed black line) modified by the smeared absorption (red), reflection from the ionized skin of the accretion disk (green) and the emission associated with the smeared wind system (blue). {\it Middle Panel} The complete best-fit model fit to PG1211+143 including both narrow ionized absorption systems, and the emission associated with this absorption, and galactic absorption. {\it Bottom Panel} The fit of this model (solid line) to the EPIC PN spectrum of PG1211+143 (upper section) and the data/model ratio residuals (lower section).}
\label{pg1211}
\end{minipage}
\end{figure}

The impact of Compton reflection is strongly dependant on the viewing angle. In the standard unified AGN picture, the accretion disk is seen almost face on in type-1 AGN such as PG1211+143, while the the region of the (putative) torus exposed to the X-ray continuum emission is almost edge-on to our line of sight. Thus we expect the total reflection signature to be dominated by the component associated with the accretion disk, which may be ionized (Januk \etal 2001), and should be relativistically smeared by the large orbital velocities. We use the XION ionized reflection model (Nayakshin \etal 2001) to represent this component. This is an extremely flexible and sophisticated model, incorporating several different geometries for the accretion disk and the X-ray source(s), variable iron abundance and a wide range of accretion rates. We choose a set of fixed values for the majority of the model parameters that are appropriate for the properties of this source. Specifically, we choose a magnetic flare based geometry for the X-ray source (which forces $\Omega$/2$\pi$=1), we fix the outer edge of the accretion disk at 1000 R$_{s}$, the height of the flares above the accretion disk at 5 R$_{s}$, the accretion rate to be 0.78\Mdotedd~(based on an estimate of L$_{bol}$/L$_{Edd}$)\footnote{L$_{bol}$=4.4$\times$10$^{45}$ erg s$^{-1}$ \& L$_{Edd}$=5.7$\times$10$^{45}$ erg s$^{-1}$ for M$_{BH, PG1211}$=4.5$\times$10$^{7}$\Msun; Kaspi \etal 2000}, the Fe abundance to be solar and the viewing angle of the disk to be almost face-on (cos($i$)=0.85). The remaining free parameters in the fit are; the radius of the innermost edge of the accretion disk (this governs the amount of relativistic broadening observed) and the ratio of the X-ray source luminosity to the luminosity of the disk (L$_{x}$/L$_{disk}$). 

On top of this base model we add the absorption and, crucially, the emission from the disk wind model discussed in Section~\ref{2}. The ionization state, column density, the velocity smearing width and the normalization of the emission are free parameters in the model. 

Figure~\ref{pg1211} shows the best-fit model continuum components (top panel), the total best-fit model (middle panel), the model fit to the EPIC PN data (bottom panel, upper section) and the data/model ratio residuals (bottom panel, lower section). Table~\ref{best-fit} details the best-fit parameters of the model. The model provides a good fit to the \xmm~EPIC PN data ($\chi^{2}_{red}$=1.08). The curvature introduced into the spectrum from the smeared absorption/emission model does an excellent job of reproducing the apparent soft X-ray excess. The only remaining residuals are narrow features around 2.2 keV (a known calibration issue around the gold edge in the EPIC PN response\footnote{Document ref: XMM-SOC-CAL-TN-0061. http://xmm.vilspa.esa.es/ external/xmm\_sw\_cal/calib/documentation/}, and $\sim$7 keV, which is probably due to the very highly ionized absorption system (Pounds \etal 2003) not being completely described by our model. We speculate that this is due to the incomplete Fe K-shell physics in the version of XSTAR used to perform the photoionization calculations here (v2.1kn3 as opposed to the newest version available, v2.1l, which includes more detail for Fe K-shell transitions: Kallman \etal 2004). 

We note that our spectra decomposition attributes the majority of the soft excess to smeared absorption and emission rather than with the reflection component. Signatures of the reflection are clearly present in the spectrum (see Section~\ref{4-4}), however the best-fit parameters of the reflection continuum are such that this component does not rise strongly below 0.7 keV. A note of caution is required here however because both reflection dominated and absorption dominated models are both capable of providing a good fit to these data. This implies that either the models (in particular the details of the ionization and velocity structure of the wind and/or the accretion disk) or the data are of insufficient quality to distinguish these two physics scenarios unambiguously (Sobolewska \& Done 2006).

\subsection{The smeared absorption/emission system}
\label{4-3}

Fig.~\ref{pg1211} shows that the apparent soft excess is well modelled by the combination of the smeared absorption and emission from a column with N$_{H}$$\sim$1.5$\times$10$^{23}$ cm$^{-2}$ and velocity smearing of $\sim$0.2c. The solid angle implied by the amount of emission is $\sim$0.4, so a substantial fraction of the central source is covered by this material, as required by the observed ubiquity of the soft excess. The pronounced dip below $\sim$0.6 keV predicted for a smeared absorption/emission system with this covering fraction is not obvious in the data, however the model provides a good fit, demonstrating that this feature is masked in the observed spectrum by the reflection component and the impact of galactic absorption. 

Comparing these best-fit parameters with those for the absorption-only model presented in GD04, highlights the important effect that the emission has on the implied physical properties of the wind. The required column is reduced by a factor 2 by including the emission (from $3.3\times 10^{23}$~cm$^{-2}$ in GD04), though its ionization state is not significantly affected (560 here {\it c.f.} 460 erg cm s$^{-1}$ in GD04). More importantly, the velocity smearing is reduced by a factor 30\% (from 0.28c in GD04). This is primarily the result of the emission component helping to fill-in the strong absorption around 0.9 keV, reducing the level of smearing required to blend the two strongest absorption regions together. Thus including the emission makes the wind significantly less extreme in column and velocity than inferred from using the absorption alone.

\subsection{Compton reflection}
\label{4-4}

In line with what we expect from Narrow-line Seyfert 1 objects, the underlying continuum emission is steep, with $\Gamma$$\sim$2.4, and the spectrum displays clear signatures attributable to Compton reflection from the accretion disk. The iron K$\alpha$ line is broader than the instrument response, but not to the level expected from emission from the last stable orbit even from a Schwarzschild black hole. The lower bound on the disk radius for the line is $26R_g$, so does not require extreme spin, and the model itself by construction has $\Omega/2\pi=1$. Both these results are, however, model dependent. These data can be equally well fit without smeared absorption/emission using a continuum which is reflection dominated and requires extreme spin (Sobolewska \& Done 2006). Plainly the origin of the soft excess has profound implications for decoding the geometry of the space-time in the vicinity of a super-massive black hole, as well as for the details of hard X-ray source(s) and the accretion disk.

\begin{table}
\centering
\begin{minipage}{85 mm} \centering
\caption{Best-fit spectral parameters}
\hspace{-15mm}
\begin{tabular}{lclc}
Parameter & Value & Parameter & Value \\ \hline
$\Gamma$ & 2.370$^{+0.003}_{-0.009}$ & L$_{x}$/L$_{disk}$ & 0.83$^{+0.45}_{-0.27}$ \\
K$_{pl}$$^{a}$ & 1.43$^{+0.17}_{-0.16}$ & r$_{inner}$$^{b}$ & 27$^{+12}_{-1}$ \\
 & & & \\
N$_{H,1}$$^{c}$ & 0.1$^{+0.01}_{-0.01}$ & N$_{H,2}$$^{c}$ & 1.00$^{+0.06}_{-0.04}$ \\
$\xi_{1}$$^{d}$ & 1.988$^{+0.002}_{-0.006}$ & $\xi_{2}$$^{d}$ & 3.27$^{+0.01}_{-0.01}$ \\
v$_{out,1}$$^{e}$ & 13.7$^{+1.5}_{-1.5}$ & v$_{out,2}$$^{e}$ & 44.9$^{+0.9}_{-0.5}$ \\
C$_{f,1}$ & 0.005$^{0.06,\,f}$ & C$_{f,2}$ & 0.002$^{0.03,\,f}$ \\
 & & & \\
N$_{H,wind}$$^{c}$ & 1.48$^{+0.03}_{-0.03}$ & $\xi_{wind}$$^{d}$ & 2.748$^{+0.021}_{-0.002}$ \\
$\sigma_{wind}$ & 0.207$^{+0.005}_{-0.004}$ & C$_{f,wind}$ & 0.4$^{+0.02}_{-0.05}$ \\
 & & & \\
 & & $\chi^{2}$ & 1023.2 \\
 & & d.o.f & 949 \\ \hline
\multicolumn{2}{l}{\scriptsize $^{a}$ 10$^{-3}$ photon keV$^{-1}$ cm$^{-2}$ s$^{-1}$} & \multicolumn{2}{l}{\scriptsize $^{d}$ erg cm s$^{-1}$} \\
\multicolumn{2}{l}{\scriptsize $^{b}$ Schwarzschild radii} & \multicolumn{2}{l}{\scriptsize $^{e}$ 10$^{3}$ km s$^{-1}$} \\
\multicolumn{2}{l}{\scriptsize $^{c}$ $\times$10$^{23}$ cm$^{-2}$} & \multicolumn{2}{l}{\scriptsize $^{f}$ 1$\sigma$ Upper Limit}\\
\end{tabular} \hspace{-1 cm}
\label{best-fit}
\end{minipage}
\end{table}

\subsection{The narrow absorption systems}
\label{4-5}

The origin of the soft excess also has implications for the narrow absorption systems. Both the smeared absorption/emission model and the smeared ionized reflection model contain an intrinsically steep continuum spectrum (typically modelled with a power-law with $\Gamma$$\sim$2.2-2.4), while interpreting the soft excess as a separate continuum component (as the tail of the multi-colour blackbody emission from the accretion disk, for example) suggests a much flatter spectrum, ($\Gamma$$\sim$1.8-1.9). The ion populations of the narrow absorption systems depend strongly on the spectral index of the ionizing continuum and will yield subtly different narrow absorption systems in each case. Both Pounds \etal (2003) and Kaspi \etal (2006) interpret the soft excess as a separate continuum component and, as a result, the properties of the narrow absorption systems fit here are different to those that Pounds \etal \& Kaspi \etal derive from the same data.

In particular, while the ionization states and column densities found here are rather similar to those of Pounds \etal (log($\xi$)$\sim$3.27 \& 1.99 {\it c.f.} $\sim$3.4 \& 1.7 and N$_{H}$$\sim$1$\times$10$^{23}$ \& 1$\times$10$^{22}$ {\it c.f.} $\sim$5$\times$10$^{23}$ \& 0.6$\times$10$^{22}$ cm$^{-2}$), the inferred outflow velocities are different (45,000 \& 13,000 km/s {\it c.f.} 24,000 \& 24,000 km/s, respectively). The steeper continuum in our model prevents a strong Fe{\small XXVI} Ly$\alpha$ (6.96 keV) transition being present in the spectrum at $\log_{10}\sim 3.3$. Instead, the spectral feature at $\sim$7 keV (observed frame) is modelled with the wealth of Fe absorption lines present in our model at a rest frame energy of $\sim$6.6 keV, resulting in a considerably higher apparent outflow velocity for the most ionized narrow absorption system. Again, we note that the outflow velocity could change using the newest K shell calculations (Kallman \etal 2004). 

None of the narrow absorption systems has significant emission detected from it. At CCD resolution the individual lines cannot be resolved, but we can place fairly stringent upper limits on the covering fraction of this material at less than $\sim$6\% of the sky. Examining the emission spectra in detail reveals that the covering fraction of the more strongly ionized gas is constrained primarilly by the lack of strong ionized iron K features around $\sim$7 keV, whereas the covering fraction of the more weakly ionized material is constrained by the lack of strong soft X-ray features, particularly those related to OVII ($\sim$0.7 keV) and OVIII ($\sim$0.9 keV).

We note that the turbulent velocity used in the photoionization modelling of the narrow absorption systems does affect the observability of the emission spectrum, however the effect is weak with velocities up to $\sim$10,000 kms$^{-1}$ resulting in a factor $<$2 change in the upper limit.

\section{Concerning disk wind properties}
\label{5}

Given our best-fit wind properties we can make a crude estimate of the inferred mass-loss rate, $\dot{M}_{out}$. Assuming that our velocity smearing represents a geometrically thick ({\it i.e.} $R_{out}$$>>$$R_{in}$) outflowing wind, the mass-loss rate approximates to;

\begin{flushright}
\begin{minipage}{80mm}
\begin{equation}{\hspace{5mm}}\dot{M}_{out}=C_{f}\,4\pi R_{in}^{2}\,n_{e}m_{p}\,v
\label{eq8}
\end{equation}
\end{minipage}
\end{flushright}

\hspace{-8mm} where $C_{f}$ is the global covering fraction of the wind, $R_{in}$ is the wind launch radius, $n_{e}$ is the electron density, $m_{p}$ is the mass of the proton and $v$ is the average velocity of the wind material. Using the parameters for our photoionized material (R$_{in}$=5$\times$10$^{14}$ cm and $n\sim$10$^{12}$ cm$^{-3}$), $\dot{M}_{out}$=190\Msun~yr$^{-1}$. By contrast, the mass accretion rate which corresponds to accretion at the Eddington limit for PG1211+143 (assuming an accretion efficiency of $\sim$6\%) is $\sim$1.3\Msun~yr$^{-1}$, presenting us with the uncomfortable situation that the disk is losing over 100 times more mass through the wind than is being accreted.

Another way to see the same conflict is via the column density, N$_{H}$, through the wind,

\begin{flushright}
\begin{minipage}{80mm}
\begin{equation}{\hspace{5mm}}N_{H}\,=\int_{R_{in}}^{R_{out}}\hspace{-4mm}n_{e}\,dR\,=\int_{R_{in}}^{R_{out}}\hspace{-4mm}\frac{\dot{M}_{out}}{4\pi R^{2}\,C_{f}\,m_{p}v}\, dR
\label{eq9}
\end{equation}
\end{minipage}
\end{flushright}

For a constant velocity outflow extending to $\infty$ then N$_{H}$=1.2$\times$10$^{27}$ cm$^{-2}$, dramatically larger than the measured column needed to produce the soft excess ($\sim$10$^{23}$ cm$^{-2}$). 

Clearly the main problems lie in the assumption of a uniform density of the material and that the wind extends to $\infty$ with a single, large velocity. Clearly a more detailed description of the density structure is required in order to evaluate the mass-loss rate from such a wind more accurately ({\it e.g.} the description given by Murray \& Chiang - 1995, see Section~\ref{5-1}), however if the wind slows down, as we might expect given that it has to escape from a large gravitational potential well, then the predicted $\dot{M}$ and N$_{H}$ increase even further!

Chevallier \etal (2005) perform a similar set of calculations for their pure- absorption, constant pressure, models (using a standard set of parameter values)  and again they calculate a mass-loss rate of $\dot{M}_{out}$$\sim$10$^{3}$ $\dot  M_{Edd}$ unless they incorporate an extremely low volume-filling factor.  Considering that we derive our wind parameters from detailed spectral fitting,  and our model includes emission as well as absorption, we consider the agreement  between our results and the Chevallier \etal calculations to be rather good.  This presents a major problem for the validity of interpreting and origin for  the smeared absorption/emission as arising in a disk wind. 

Examining the model in detail a possible method of circumventing much of this problem suggests itself. As noted in Section~\ref{2-2}, the amount of turbulent velocity present in the material strongly affects the opacity of heavily saturated absorption lines and analysing such a spectrum with a model that only incorporates a small turbulent velocity will result in a considerable overestimate of the column density of the material. Simulated XSTAR spectra with turbulent velocities of $\sim$3$\times$10$^{4}$ km s$^{-1}$ (the maximum allowed in the code) show that this effect can result in up to $\sim$3-4 times more absorption in the 0.3-3 keV band, for gas with similar properties to the gas we use in our simulations, than for gas with small turbulent velocities.

Fortuitously, whilst the velocity dispersion an accretion disk wind will differ considerably from that imposed by turbulent velocity, the extremely simple Gaussian smearing used in this work is actually rather similar to the velocity dispersion from turbulent velocity (although, obviously, it does not incorporate the increase in opacity for heavily saturated lines). The required velocity shear we apply ($\sim$6$\times$10$^{4}$ km s$^{-1}$), if interpreted as a turbulent velocity instead of a wind outflow velocity, might well be causing us to overestimate the column density of our material, and hence out mass-loss rate, by up to a factor $\sim$10. Clearly this correction is not capable of producing the three orders of magnitude required to solve the mass-loss rate problem.
 
\subsection{Comparison with simulations}
\label{5-1}

Detailed hydrodynamic simulations are a relatively new addition to the study of accretion disk winds, however a comparison between the spatial simulations presented in PSK00 \& PK04, and spectral simulations presented here, is instructive. The basic properties of the spatial simulations are relatively well matched to PG1211+143, with only the ratio L$_{x}$/L$_{disk}$ and the spectral index of the ionizing X-ray radiation being significantly different (see Table~\ref{pg1211vsim}).

\begin{table}
\centering
\begin{minipage}{85 mm} \centering
\caption{Comparing the physical parameters of PG1211+143 with those of accretion disk wind simulations (PK04)}
\hspace{-15mm}
\begin{tabular}{lcc}
 & PG1211+143 & Simulations \\ \hline
$M_{BH}$(\Msun) & 4.5$\times$10$^{7}$ & 10$^{8}$ \\
$\dot M$/\Mdotedd & 0.78 & 1 \\
(L$_{x}$/L$_{disk}$) & 0.8 & 0.1 \\
$\Gamma_{X-ray}$ & 2.4 & 1.7 \\ \hline
\end{tabular} \hspace{-1 cm}
\label{pg1211vsim}
\end{minipage}
\end{table}

The spatial disk wind simulations show a wind with an extremely complex density and velocity structure, clearly showing that the constant density, thin shell, treatment assumed in our photoionization calculations is far too simplistic. Not only is the density structure complex, but the typical density of the the simulated wind is much lower than the value assumed during our modelling (10$^{9}$ cm$^{-3}$ {\it c.f.} 10$^{12}$ cm$^{-3}$), except in the very innermost regions. We note, however, that the harder spectral index of the spatial simulations will result in a more strongly ionized wind, driving down the density of the wind in comparison with the values expected from a considerably softer X-ray source. The spatial simulations also show that the global wind velocity structure is extremely complex and strongly time dependant, in strong contrast with the simplistic velocity treatments discussed in Section~\ref{2-2}. 

The region of the spatial simulations that best matches our ionization and column density conditions (log$_{10}$($\xi$)$\sim$2-4 and/or a column of $N_{H}$$\sim$10$^{23}$ cm$^{-2}$) is the fast, relatively geometrically thin, high-density `stream' which, in addition to having approximately the correct ionization state and column density, also displays the required large velocity dispersion {\it and} peaks at speeds of $\sim$30,000 km s$^{-1}$ (this is, in fact, a lower limit on the wind velocity since there is a numerical ceiling on the velocities allowed in the simulations). We note that despite this apparent match, for the particular simulation shown in PK04, there is {\it no} specific l.o.s through the fast stream which has {\it both} the observed column and the correct ionization state simultaneously. While the global wind velocity structure in the spatial simulations is extremely complex and strongly time dependant, making the correct velocity smearing for the emission difficult to calculate, absorption in the material in the fast stream is a simpler case. 

Murray \etal (1995) and Murray \& Chiang (1995) describe the velocity and density structure of an accretion disk wind along a single streamline. The fast stream in the spatial simulations is extremely similar to this case, allowing us to combine the Murray \& Chiang descriptions of the density and velocity distribution along a streamline into a single density-weighted radial velocity smearing function (which turns out to be proportional to R$^{-2}$) for the fast stream. Applying this smearing function in future work will be a considerable improvement in the treatment of absorption by accretion disk winds in AGN and, provided that the emission from this fast stream dominants the emission from the wind material as a whole, may also be appropriate for the absorption. A more fundamental problem here is that many of the PG Quasars show soft X-ray excesses that are very similar to that of PG1211+143. In the interpretation as a disk wind, this implies that all these sources see through a similar region of the wind, exhibiting a similar amount of absorption in material with a similar ionization state. While the covering fraction of the fast stream is of order $\sim$0.2 (PSK00, PK04), the section which closely matches our ionization and column density conditions is much smaller. Unfortunately, while the fast stream of a line-driven disk wind may have some of the properties required to reproduce the soft X-ray excess in PG1211+143, it is unlikely that the l.o.s to so many PG Quasars would fall within the tightly constrained range of solid angle covered by the fast stream of such a wind. The situation becomes even more problematic when we consider that all these objects are Type-1 AGN, implying that the l.o.s is close to the normal to the plane of the disk. In contrast, the fast stream in the disk wind simulations is generally quite equatorial, rising no more than 25$^\circ$ from the disk plane. The simulations of line-driven winds seem to bear a striking resemblance to the outflows required in broad absorption line quasars (PSK00, PK04), but not to the rather faster, denser and larger scale height material required here.

In addition, the total mass-loss rate in the spatial simulations is only 0.5\Msun~yr$^{-1}$, in stark contrast to the 190\Msun~yr$^{-1}$ predicted by out spectral model. At first glance, it seems that the huge mass outflow rates implied by our spectral modelling can be simply reduced by dropping the density used in the photoionization calculations to a density more typical of that seen in the wind simulations. Reducing the density used in the spectral simulations to $\sim$10$^{9}$ cm$^{-3}$ would result in a reduction of the mass loss rate to $\dot M_{outflow}$$\sim$0.1-1\Msun~yr$^{-1}$. We then require, however, that the material is distributed in a thick shell, with $\Delta$R$>>$R in order to obtain the required column density of $\sim$10$^{23}$ cm$^{-2}$. Unfortunately, relaxing the `thin shell' condition generates its own obstacles, namely {\it i}) a considerable increase the computational time required to calculate the grid models, and {\it ii}) photon-loss, and hence energy-loss, through the upper/lower `surface' of the wind (see Appendix~\ref{app-2} \& King \& Pounds !2003 for a brief description of this problem).

Once again, however, we note that the more strongly ionized wind that results from the harder spectral index of the spatial simulations, not only drives down the density of the wind, but also leads to less material reaching escape velocity before becoming over-ionized. As the material becomes increasingly more ionized, it becomes transparent to the UV line emission (as there are few ion stages left with substantial line opacity), reducing the mass-loss rate from such a wind in comparison with a wind exposed to a softer X-ray continuum. 

In fact, a sufficiently powerful hard X-ray flux causes the wind to become so over-ionized that the fast stream produced by disk dominated spectra gives way to a `failed wind' structure (Proga 2005). In this case, the material rises above the disk due to UV absorption and starts to accelerate, but as it is exposed to the X-ray source it becomes over-ionized, thus preventing continued acceleration, and the material eventually falls back down to the disk rather than escaping as a wind. This situation can result in quite high densities at large scale heights above the disk, complete with a complex, circulating velocity field and, crucially, very little mass-loss (Proga 2005). Since this occurs directly above the inner disk, the velocity dispersion is expected to be large and it is possible that magnetic fields may help to stir the material up to even greater scale heights, increasing the covering fraction of this material.

\section{Conclusion}
\label{6}

GD04 proposed a model for the soft excess, observed in the X-ray spectra of many type-1 AGN, based on heavily smeared, ionized, absorption. They suggested a tentative origin for this absorption in the innermost, accelerating, region of an accretion disk wind. We expand on this model by including the emission that {\it must} be associated with this absorption, focusing on the impact that the emission has on the total spectrum. Initially, we examine the properties of the emission, the absorption and the total spectrum as we vary the free parameters in the model, before testing the model against data for PG1211+143, a bright nearby quasar known to have one of the strongest soft X-ray excesses. 

The main conclusion from this work is that the emission associated with heavily smeared, ionized, absorption actually helps this model reproduce the soft excess, by filling-in some of the absorption trough. This removes some of the fine tuning inherent in the original pure absorption model and, rather than reducing the curvature imposed on the spectrum by the model, the emission instead allows the model to reproduce the soft excess curvature between 0.6-1.5 keV from a wider range of column densities and ionization states, with lower velocity smearing. 

We show this specifically by fitting the total smeared absorption/emission to the \xmm~EPIC PN spectrum of PG1211+143, as part of a more complex model that accounts for complex narrow absorption systems and Compton reflection from an ionized accretion disk. This model results in a good fit to the data ($\chi^{2}_{red}$=1.08) demonstrating that it is capable of reproducing even the strongest soft X-ray excesses observed. Furthermore, in comparison with the details discussed in GD04, the addition of the emission reduces the column density required to fit these data by a factor $\sim$2 and reduces the smearing velocity from $\sim$0.28c to $\sim$0.2c. The accretion disk reflection component does contribute to the spectrum, but it does not produce a strong contribution to the soft excess. The reflection is smeared, but not dramatically so. This is in sharp contrast to the reflection dominated models for the soft excesses, which require extreme spin of the black hole. 

However, there are significant problems with interpreting the absorbing material as an accretion disk wind. The interpretation as a UV line-driven disc wind, suggested in GD04, is especially problematic because such winds tend to be produced within $\sim$25$^\circ$ of the equatorial plane (PK04). Thus most type-1 AGN are expected to have lines of sight which would not intercept much of this material. This directly conflicts with the observations that most PG Quasars (type-1 AGN) show a soft excess. There is also a more generic problem in identifying this material as a wind. The required large velocity smearing implies a mass-loss rate of $\dot M_{outflow}\sim$190\Msun~yr$^{-1}$, much larger than the accretion rate required to power the observed luminosity of $\sim$1-2\Msun~yr$^{-1}$ (see also Chevallier \etal 2005). 

We suggest an origin for this material as a `failed wind', {\it i.e.}, that the central X-ray source is strong enough to over-ionize the wind, removing the acceleration through line absorption before the material reaches escape velocity. Thus the material is initially accelerated, but it cannot escape the system and eventually falls back to the disk. The unreasonably large mass-loss rate is then easily circumvented, since most of the material simply circulates, falling back and rejoining the disk at somewhat larger radii. Modelling such material is undoubtedly much more complex than described by our very simplistic assumptions on the density and velocity structure, but despite this we get a very good fit to the shape of the soft excess in PG1211+143 from a combination of absorption and emission in this material. Better models fit to a wide range of type-1 AGN will allow future studies to determine if this interpretation is physically plausible.

\section{Acknowledgements}

The authors thank Tim Kallman for writing and supporting the XSTAR code, Daniel Proga, for giving us the results from his hydrodynamic simulations, and  Marek Gierli\'nski, Chris Simpson \& Anthony Brown for useful discussion and  comments. This work is based on observations obtained with \xmm, an ESA science  mission with instruments and contributions directly funded by ESA Member States  and the USA (NASA). This research has made extensive use of NASA's Astrophysics  Data System Abstract Service.  NJS and CD acknowledge financial support through a PPARC PDRA and Senior  fellowship, respectively.

\appendix

\section{Details of the emitted luminosity}
\label{app-1}

The equations used for photoionization codes ({\it e.g.} CLOUDY, XSTAR, TITAN, etc) are complex and are presented in detail elsewhere (most notably in the manuals for the specific codes). However it is instructive to understand some of the dependencies of bound-bound (lines), bound-free (RRC) and free-free (ff) interactions in a photoionized gas.

Equations~\ref{eqa1}, \ref{eqa2} \&~\ref{eqa3} are expanded versions of equation~\ref{eq2} (Section~\ref{2}), showing some the complexities of the $F_{n}$-functions, We note that even these equations have been considerably simplified, for example we ignore collisional ionization and three body interactions. Each of these equations basically contains a sum of the energy released from the probable interactions for each ion species, $i$, in the gas, weighted by the fraction of the gas made up by this ion, $f_{i}$ (Rybicki \& Lightman 1979, Osterbrock 1989). 

\subsection{Radiative Recombination}
\label{app1-1}

The luminosity emitted by radiative recombination transitions is given by,

\begin{flushright}
\begin{minipage}{80mm}
\begin{equation}
\hspace{-4mm}L_{RRC}\,=\,n_{e}^{2}\,V\sum_{i}f_{i}\left(\sum_{x}\hspace{-2mm}\alpha_{RRC,i,x}\left[\int_{\nu_{min}}^{\infty}\hspace{-4mm}h\nu_{i}(x)\,d\nu_{i}\right]\right)
\label{eqa1}
\end{equation}
\end{minipage}
\end{flushright}

\hspace{-6mm}Again, $n_{e}$ is the electron density of the material, $f_{i}$ is the fraction of the gas in the specific ion species, $i$. $\alpha_{RRC,i,x}$ is the radiative recombination coefficient for the recombination of a free electron to a specific energy level, $x$, in the ion species, $i$. The integral is the total energy released by recombinations into energy level, $x$, for ion, $i$, by the total distribution of free electrons, where $\nu_{min}$ is the minimum frequency of a photon released by the recombination of a barely-free electron to energy level, $x$.

\subsection{Line emission}
\label{app-1-2}

Since line emission in a photoionized gas relies (to first order) on a previous radiative recombination in order for a line emitting cascade to occur, the luminosity emitted in line transitions is given by,

\begin{flushright}
\begin{minipage}{80mm}
\begin{equation}
\hspace{-4mm}L_{lines}\,=\,n_{e}^{2}\,V\,\sum_{i}f_{i}\,\alpha_{RRC,i}\left(\sum_{t}\,P_{i,t}\,h\nu_{i,t}\right)
\label{eqa2}
\end{equation}
\end{minipage}
\end{flushright}

\hspace{-6mm}$n_{e}$ is the electron density of the material, $f_{i}$ is the fraction of the gas in the specific ion species, $i$, and $\alpha_{RRC,i}$ is the radiative recombination coefficient for ion species, $i$. $P_{i,t}$ is the probability of a particular {\it set} of atomic energy level transitions, $t$, occurring for the specific ion species, $i$, and $h\nu_{i,t}$ is the total energy of the photons released by that particular set of atomic level transitions. Here $\alpha_{RRC,i}$ is given by,

\begin{flushright}
\begin{minipage}{80mm}
\begin{equation}
\alpha_{RRC,i}\,=\,\sum_{x}\alpha_{RRC,i,x}
\label{eqa1-sum}
\end{equation}
\end{minipage}
\end{flushright}

\hspace{-6mm}We note that the $n_{e}^{2}$ dependence of this component is forced on it by the requirement that a 2-body (electron-ion) recombination event occurs prior to the line emission.

\subsection{Free-free emission}
\label{app-1-3}

The luminosity emitted by free-free integrations is given by,

\begin{flushright}
\begin{minipage}{80mm}
\begin{equation}
\hspace{-4mm}L_{ff}\,=\,n_{e}^{2}\,V\,\sum_{i}f_{i}\,\alpha_{ff,i}\left(\int\,h\nu_{i}\,d\nu_{i}\right)
\label{eqa3}
\end{equation}
\end{minipage}
\end{flushright}

\hspace{-6mm}$n_{e}$ is the electron density of the material, $f_{i}$ is the fraction of the gas in the specific ion species, $i$. $\alpha_{ff,i}$ is the free-free coefficient for ion species $i$ and the integral is the total energy released by the free-free integrations of ion, $i$, with the total distribution of free electrons.

\section{Photon loss through the surfaces of a thick wind}
\label{app-2}

Consider a simple `toy' model of an accretion disk wind as uniform outflowing gas in a conic section torus with a with an opening angle, $\phi$, and a well-defined upper, lower and outer boundary. Photons entering the material are reprocessed through absorption, re-emission and scattering and, to first order, reprocessed photons are distributed isotropically with each interaction. Photons produced by interactions near the upper/lower surface of the wind are capable of escaping from the wind through these surfaces, significantly changing the amount of energy available for following interactions in the remaining material (King \& Pounds 2003). 

If the wind is radially thin this effect is negligible, given the small surface area available for photons to be lost through, and so a thin shell (no upper/lower surfaces) is a reasonable approximation of a thin wind. This is the case presented in the model here. This is {\it not} the case for a thick wind which, with its considerably larger surface area, will not be well approximated by a thick shell of gas. A thick wind can be crudely approximated with XSTAR by allowing the {\it local} covering fraction to be, {\it cfrac}$<$1. This allows photons to escape within each calculational sub-layer in the material, however this condition then applies throughout the shell of material, not just to the surface layers.

\end{document}